\documentclass[aps,nofootinbib,amsfonts,superscriptaddress]{revtex4}

\usepackage{amsfonts,amssymb,amscd,amsmath}

\usepackage{graphicx}
\usepackage{subfigure}
\usepackage{color}
\usepackage{soul}
\DeclareGraphicsExtensions{.pdf,.jpg,.png,.gif}

\graphicspath{{figures/}}
\usepackage[utf8]{inputenc}

\newcommand{\beql}[1]{% Begin enumerated equation with label #1
  \begin{equation}\label{eq:#1}}
\newcommand{\eeq}{% End equation mode
  \end{equation}}

\definecolor{brightpink}{rgb}{1.0, 0.0, 0.5} % Colour for TB comments.

\begin{document}

%\title{Machine Learning enhanced Multi-messenger Probes for New Physics and Cosmology}

\title{Machine Learning for Multi-messenger Probes of New Physics and Cosmology: A Review and Perspective}

\author{Andrea Addazi}
\affiliation{Center for Theoretical Physics, College of Physics Science and Technology, Sichuan University, 610065 Chengdu, China}
\affiliation{INFN, Laboratori Nazionali di Frascati, Via E. Fermi 54, I-00044 Roma, Italy, EU}

\author{Konstantin Belotsky}
\affiliation{National Research Nuclear University “MEPHI”, 115409 Moscow, Russia}
\author{Vitaly Beylin}
\affiliation{Virtual Institute of Astroparticle Physics, 75018, Paris, France, EU}

\author{Timur Bikbaev}
\affiliation{National Research Nuclear University “MEPHI”, 115409 Moscow, Russia}
\affiliation{Research Institute of Physics, Southern Federal University, 344090 Rostov on Don, Russia}

\author{Deen Chen}
\affiliation{College of Design and Innovation, Tongji University, 281 Fuxin Rd, 200092 Shanghai, China}

\author{Filippo Fabrocini}
\affiliation{College of Design and Innovation, Tongji University, 281 Fuxin Rd, 200092 Shanghai, China}
\affiliation{Institute for Computing Applications “Mario Picone”, Italy National Research Council, Via dei Taurini, 19, 00185 Rome, Italy, EU}

\author{Stefano Giagu}
\affiliation{Department of Physics, University of Rome “La Sapienza”, Piazzale Aldo Moro,
5, 00185 Rome, Italy, EU}
\affiliation{Istituto Nazionale Fisica Nucleare, Sezione di Roma, Piazzale Aldo Moro,
5, 00185 Rome, Italy, EU}

\author{Krid Jinklub}
\affiliation{College of Design and Innovation, Tongji University, 281 Fuxin Rd, 200092 Shanghai, China}
\affiliation{College of Arts, Media, and Technology, Chiang Mai University, 239 Huay Kaew Rd, 50200, Chiang Mai, Thailand}

\author{Artem Kharakhashyan}
\affiliation{Research Institute of Physics, Southern Federal University, 344090 Rostov on Don, Russia}

\author{Maxim Khlopov}
\affiliation{Virtual Institute of Astroparticle Physics, 75018, Paris, France, EU}

\author{Vladimir Korchagin}
\affiliation{Research Institute of Physics, Southern Federal University, 344090 Rostov on Don, Russia}

\author{Maxim Krasnov}
\affiliation{National Research Nuclear University “MEPHI”, 115409 Moscow, Russia}

\affiliation{Research Institute of Physics, Southern Federal University, 344090 Rostov on Don, Russia}

\author{Atharv Mahajan}
\affiliation{International Centre for Space and Cosmology, Ahmedabad University, Ahmedabad 380009, India}

\author{Antonino Marcian\`o}
\email{marciano@fudan.edu.cn}
\affiliation{Center for Field Theory and Particle Physics $\&$ Department of Physics, Fudan University, 200433 Shanghai, China}
\affiliation{Laboratori Nazionali di Frascati INFN,
Via Enrico Fermi, 54, 00044 Frascati (Rome), Italy, EU}
\affiliation{INFN sezione Roma Tor Vergata, I-00133 Rome, Italy, EU}

\author{Andrey Mayorov} 
\affiliation{National Research Nuclear University “MEPHI”, 115409 Moscow, Russia}
\author{Antonio Morais}
\affiliation{Laboratório de Instrumentação e Física Experimental de Partículas (LIP), Universidade do Minho,
4710-057 Braga, Portugal, EU}

\affiliation{Departamento de Física, Escola de Ciências, Universidade do Minho, 4710-057 Braga, Portugal, EU}

\author{Roman Pasechnik}
\affiliation{Department of Physics, Lund University, S\"olvegatan 14A S 223 62 Lund, Sweden, EU}

\author{Jackson Levi Said}
\affiliation{Institute of Space Sciences and Astronomy, University of Malta, Malta, MSD 2080} 
\affiliation{Department of Physics, University of Malta, Malta, EU}

\author{Danila Sopin}
\affiliation{National Research Nuclear University “MEPHI”, 115409 Moscow, Russia}

\affiliation{Research Institute of Physics, Southern Federal University, 344090 Rostov on Don, Russia}
\author{Viktor Stasenko}
\affiliation{National Research Nuclear University “MEPHI”, 115409 Moscow, Russia}

\author{Oem Trivedi}
\affiliation{Department of Physics and Astronomy, Vanderbilt University, Nashville, TN 37235, USA}

%\date{November 2022}

%\maketitle

\begin{abstract}
\vspace{0.5cm}
\noindent
The multi-messenger exploration of dark matter and physics beyond the Standard Model has emerged as a central direction in modern astro-particle physics, particularly following the discovery of gravitational waves. In this work, we present a comprehensive review and forward-looking perspective on machine-learning-enhanced multi-messenger approaches, combining information from gravitational waves, cosmic rays, gamma rays, neutrinos, and collider experiments. We summarize the current state of the field, discuss recent methodological developments, and outline a coherent research program aimed at integrating heterogeneous datasets within a unified inference framework. Our collaboration proposes here a plan for forthcoming analyses aiming at extracting information on the properties and interactions of dark matter, and finally on its genesis, combining multi-messenger astronomy techniques and inputs from laboratory physics. The main objectives planned in this line of research comprise: i) the multi-messenger analysis of new physics in cosmology, including mainly, but not only, several different models of dark matter; ii) the phenomenology of new physics signatures in ground-based cosmic rays experiments, with cross-correlation to the corresponding physical, astrophysical and cosmological observations; iii) the development of machine learning methods for data analysis in ground-based cosmic rays experiments, in light of the new physics signatures. We note that several groups have explored the use of multi-messenger observations, including gravitational waves, to probe alternative dark matter candidates. The present work builds on these developments by focusing on the role of machine learning in integrating heterogeneous datasets. We foresee that a cross-fertilizing approach combining the information that arise from so different experimental methodologies will represent the right and successful path to extract information about the very elusive dark matter particles and provide answers to the main questions that are left in fundamental physics.
\end{abstract}

\maketitle

\tableofcontents

\pagebreak

\section{Introduction and general State-of-the-Art}
\noindent
One of the most pressing questions in modern physics concerns the nature of dark matter (DM) and its interplay with gravitational and cosmological observations. Addressing this problem requires combining information from multiple observational channels, including electromagnetic radiation (e.g. CMB measurements by Planck, ACT, Ali-CPT) \cite{Planck2018Overview,ACT2020DR4, AliCPT_ref}, cosmic rays and neutrinos (LHAASO, HAWC, HESS, IceCube) \cite{LHAASO, HAWC2017Overview, HESS2006Instrument, IceCube}, collider experiments (LHC, HL-LHC, CEPC) \cite{LHC2008Machine,HILHC_ref, CEPC_ref}, and gravitational-wave observations (LIGO--Virgo, LISA, DECIGO, BBO) \cite{LIGO, LISA:2017pwj, UDECIGO_ref, BBO_ref}, as well as pulsar timing arrays (FAST, SKA, IPTA) \cite{FAST_ref, SKA_ref, IPTA_ref}.
\\

The emergence of multi-messenger astrophysics has opened a qualitatively new avenue for probing dark matter and physics beyond the Standard Model. By combining heterogeneous datasets across vastly different energy scales and detection techniques, multi-messenger analyses can significantly enhance sensitivity to non-standard scenarios, particularly those that are difficult to access within single-channel approaches.
\\

Despite rapid progress in recent years, the field remains fragmented: existing studies typically focus on individual observables or limited combinations of datasets, and a systematic framework for consistently integrating multi-messenger information is still lacking. 

\subsection{Experiments for the detection of cosmic rays and multi-messenger probes for new physics and cosmology}
\noindent
We discuss here a novel approach to the  multi-messenger analysis of DM models beyond the WIMP miracle paradigm, investigating several possible models provided by the theoretical literature, while resorting to a sharply developed methodology that accounts for machine learning techniques hitherto applied in the literature. The observables for our multi-messenger analysis will be primarily provided by data on cosmic rays spectra, electromagnetic radiation (gamma rays) and on GW that are about to be released either by direct measurements through gravitational interferometers or indirect measurements by radio-telescopes. Indeed, within this vast and puzzling scenario, the direct GW detection, which was finally attained only few years ago, represents an unprecedented possibility to unveil the micro-physics of DM, providing through current and forthcoming experiments a sizeable amount of information that can be deployed to solve at least some of the current shortcomings. At the same time, data on cosmic rays spectra that will arise from very sensitive experiments like LHAASO \cite{DiSciascio:2016rgi,Bai:2019khm} will enable to explore energy ranges from 100 GeV to 100 PeV, thus providing further unprecedented possibilities to test these models.\\ 

We will then discuss here the possibility to carry out multi-messenger numerical analyses of DM candidates that are both either minimal extensions of the Standard Model (SM) of particle physics or arise as more convoluted theoretical frameworks. Among the first examples, we will focus on axion like particles, the dark photon, the dark atoms, several possible instantiations of majoron models. Among the latter models, we will consider supergravity candidates and models that emerge from unification attempts of the fundamental forces. Our purpose will be then to derive tighter constraints than the ones hitherto presented in the literature on the aforementioned models, through their comparison with the sensitivities of GW interferometers and radio telescopes, with the spectrum of primary and secondary gamma rays from the DM candidates decays (satellite experiments), with the possible signatures of Higgs or Higgs-like portals within the same decays (CEPC and collider experiments like Fermi-LAT), with the refined measurements on the spectra of cosmic rays (low energies in PAMELA and AMS-02, and high energies in LHAASO), on which we will specifically focus. 

\subsection{State of the Art of Very-High-Energy-Cosmic-Rays (VHECR) Experiments}
\noindent 
The discovery, by the IceCube collaboration, of very high energy (VHE) astrophysical neutrinos \cite{IceCube}, and then the very first direct detection of GW by the LIGO-Virgo collaboration \cite{LIGO}, marked the dawn of the new field of multi-messenger astronomy, opening a new window to explore the Universe. The experimental synergy recently disclosed by successful ongoing experiments, which is provided by the existence of multiple channels of detection, including the electromagnetic radiation, gravitational radiation, cosmic rays and neutrinos, and which is complemented by the data provided by ground based colliders, is supposed to boost progresses in the understanding of some of the most urgent questions that pertain fundamental physics, and to allow the investigation of several aspects of astro-particle physics and cosmology. Within this framework the Large High Altitude Air Shower Observatory (LHAASO) experiment, built at 4410 meters of altitude in the Sichuan province of China, stands as a high sensitivity TeV gamma-ray telescope characterised by a wide field of view and high duty cycle. Other notable experiments that will shed light on the physics of cosmic rays include HAWK and HESS.
 \\
For instance, the LHAASO experiment is currently measuring, with unprecedented sensitivity, the spectrum, the composition and the anisotropy of cosmic rays, spanning within a range of energies between $10^{12}$ eV and $10^{18}$ eV. Furthermore, it is providing a wide aperture (one stereo-radiant), continuously-operated gamma ray telescope that is able to probe the energy range between $10^{11}$ eV and $10^{15}$ eV. LHAASO will be continuously surveying the TeV sky for steady and transient sources, within the energy range between 100 GeV and 1 PeV. This allows the direct observation of high energy cosmic ray sources in the range of 100-1000 TeV. While addressing observables of different type, including electronic, muonic and Cherenkov/fluorescence components, LHASSO and the novel generation of cosmic rays experiments will enable the investigation, with very accurate resolution, of the origin, acceleration and propagation of radiation. This will be allowed by detailed measurements of the energy spectrum, composition and anisotropy of the cosmic rays. \\

The data that have been currently taken and will be further collected, are already providing a boost for the studies on active galactic nuclei (AGN), especially on blazars, the majority of AGN detected by gamma-ray telescopes \cite{Madejski2016,Ajello2022_4LAC}.  Through its enhanced flux sensitivity and the wide field of view accessible by the experiment, which are unprecedented especially when compared to other Cherenkov Telescope facilities that are currently acquiring data, a very promising possibility for the search of clear evidences of very high energy cosmic rays emitted by blazars is expected to be provided. In particular, the search for hard spectra, with energies larger than 10 TeV and emitted either from the extreme blazars, such as 1ES 0229+200, or from nearby blazars, such as Mrk 421, and the search for TeV photons from distant blazars with redshift $z\sim 1$ will provide information to reconstruct cosmic TeV background and the luminosity function of TeV blazars. This furnishes an outstanding possibility to understand the origin of ultra-high energy cosmic rays (UHECR) and PeV neutrinos, and thus at the same time to probe not only the fate of Lorentz symmetries, assessing either their violation or their deformation, but also to unveil the nature of DM by testing possible axion-like particles (ALPs). Furthermore, a plethora of equally unprecedented and rich amount of information will be provided for the understanding of relativistic jet physics, such as high-energy radiation mechanisms and acceleration of particles, and the determination of extragalactic background light (EBL). \\

For completeness, we note that each of the experimental facilities mentioned above has an extensive dedicated literature; representative references are provided in the bibliography.

\subsection{Perspective and scope.}
While multi-messenger analyses and machine learning techniques have both been widely explored, a key limitation of the current state of the field is the absence of a unified inference framework capable of consistently integrating heterogeneous datasets across different observational channels. Existing approaches typically rely on channel-specific analyses or loosely coupled combinations of observables, which limits their ability to extract robust and globally consistent constraints on dark matter models.\\

The central perspective of this work is that machine learning can provide the missing layer of integration, enabling the construction of joint inference pipelines that operate directly on multi-modal data. In this sense, the goal is not merely to apply machine learning to individual observables, but to develop architectures and methodologies that allow a coherent exploration of theory space across multiple messengers within a single framework.\\

In particular, we identify a key limitation of current approaches, namely the lack of unified frameworks capable of consistently integrating heterogeneous multi-messenger datasets. We argue that machine learning provides a natural pathway to address this challenge, enabling joint inference across observational channels and opening the possibility of coherent, high-dimensional exploration of dark matter parameter spaces.\\

Therefore, this manuscript is intended as a comprehensive review and forward-looking perspective. Rather than presenting new quantitative constraints or experimental results, our goal is to systematize existing approaches and outline a coherent research program for multi-messenger analyses of dark matter and beyond-Standard-Model physics using machine learning techniques.\\

\section{Models of Dark Matter}\label{sec2}
\subsection{Dark Matter from Supergravity}
\noindent
The lack of positive evidences for supersymmetry (SUSY) at the LHC energies may reflect the extreme case that the SUSY scale is very high. Then collider searches for SUSY particles would become impossible, but the advantage of such models is the possibility of unifying them with gravity in the framework of supergravity models. Such models can predict stable superheavy gravitino as DM candidate as well as providing physical basis for Starobinsky supergravity, reproducing Starobinsky inflation scenario on supergravity basis. Probes for such models are related to their cosmological consequences that can lead to observable effects, like Primordial Black holes (PBH) and effects of their evaporation or superheavy DM effects in high-energy cosmic rays.

In the minimal Starobinsky-Polonyi $\mathcal{N}=1$ Supergravity, superheavy gravitino are produced in decays of inflaton and Polonyi fields. This framework links the physics of inflation to the physics of superheavy gravitino DM in these models \cite{we}. The price for such unification of all the four fundamental forces and unified description of inflation and DM is the loss of supersymmetric solutions for the SM problems of divergence of the Higgs boson mass and origin of the electroweak energy scale, as well as the lack of direct experimental probes for DM particles. This makes indirect effects of supermassive DM crucial for Supergravity probes, thus making the search for such effects in the LHAASO, HAWK, HESS experiments challenging.

\subsection{Axion-like particles}
\noindent
Axion-like-particles (ALPs) comprise a wide class of light pseudo-Goldstone-bosons (PGB) that arise from several different theoretical scenarios. These particles, primarily originated by the spontaneous symmetry breaking of global symmetries beyond Standard Model (BSM), include QCD axions \cite{a1,a2,a3,a4,a5,a6}, familons and arions \cite{Wiarion, BeKhloarion}, majorons \cite{ChiMoPe, GelRo} and baryo-majorons \cite{Zubm}. As PGBs, ALPs retain interactions with the SM particles that are suppressed by energy scales regulating their spontaneous symmetry breaking, and individuating their decay constants, usually denoted with $f$. String phenomenology may also provide a rich variety of ALPs candidates beyond QCD axion mass and couplings \cite{WiPQ,ConPQ,SPQ} that arise, for instance, as moduli scalar fields either from the compactifications of Calabi-Yau manifolds, or even from anomalous Peccei-Quinn symmetries resulting from the intersection of D-branes --- the violation of the Peccei-Quinn symmetry are usually induced by world-sheet, brane, new gauge and gravitational instantons.\\

Within this wide scenario, it is worth mentioning that QCD instantons were originally proposed as a way to solve the strong CP problem, instantiating a shift of the $\theta$-parameter in the topological part of the QCD Lagrangian. Thus the QCD axion can be understood in terms of the breakdown of the U$(1)_{\rm PQ}$ symmetry, induced by the triangle-anomaly involving gluons. A periodic potential for the axion particles is hence generated by the non-perturbative topological fluctuation of the gluon fields, with a minimum being reached by the axion field at the decay constant value $f$ \cite{ax1,ax2,ax3,ax4}. This effect realises a screening of the CP violating $\theta$-term. Furthermore, anomalous triangular diagrams, as the ones induced by the interaction of the electromagnetic Pontryagin density with the axion field, may generate ALPs-photons, thus opening a pathway to ALPs detection. Photo-Pontryagin-axion interaction terms are hence responsible for a photon-ALPs mixing as source by a magnetic field background. This is the Primakoff effect \cite{Di, RaSto}, which also motivated searches for ALPs in laboratory experiments with high magnetic fields --- see e.g. the CAST experiment at CERN \cite{CAST}.\\

From a phenomenological perspective, cosmic magnetic fields along the sight line might trigger ALPs oscillations, which can be then observed thanks to VHE gamma ray sources \cite{DARM, DAMPR, MiMo, SHS, MRS}. The distance of sources may then act as an amplifier parameter. Among such sources stand blazars and specific AGN characterised by gamma beams that are pointing toward our line of sight. Blazars are among the most energetic (and distant) gamma-ray sources hitherto observed. Pair production and the interaction with the radiation backgrounds affect the mean free path for VHE photons above 100 GeV. In presence of extragalactic background light (EBL), these effects induce indeed a non-negligible probability for VHE gammas to annihilate into electron-positron pairs \cite{Ni, FaSte, Aha}. The average path length hence decreases with the energy while effects due to EBL become dominant. On the other hand, for energies below 100 GeV, the pair production effect is negligible, being the average path length comparable with the Hubble radius. We may further consider photon-ALP-photon oscillations occurring in the intergalactic space before reaching the terrestrial detectors, affecting the estimated flux in a competitive way to the electron-positron pair production \cite{DARM, DAMPR, MiMo}. Finally, photon-ALP conversion happening inside blazars shall be also taken into account. The conversion into photons may hence occur even inside our galaxy \cite{SHS, MRS}. This effect would then provide an even larger average path length of VHE gammas than expected. This would provide the picture of a more transparent Universe at high energies. Remarkably, if the hardening of the blazar spectra, in turn controlled by the coupling with ALPs, would be observed, this would provide a hint for photon-ALPs oscillation in the 1$\div$100 TeV range of energies.\\

It is also worth mentioning that axion-like particle could be considered as a source of baryon asymmetry within spontaneous baryogenesis model. Arising as an angular degree of freedom of complex scalar field after symmetry breaking, axion-like particle is derivatively coupled to the fermion current, which causes baryon and lepton numbers non-conservation, while it oscillates. Such models are considered in refs. \cite{Dolgov_1995, Dolgov_1997, Cohen:1988kt}. It is demonstrated in aforementioned papers that in flat space-time baryon asymmetry in this model in case of small oscillations is proportional to the cube of the initial phase of the axion-like particle: $(n_b-n_{\overline{b}}) \propto \theta_i^3$. Due to mathematical complexity, there is no analytical result for baryon asymmetry in case of expanding space-time and it could be studied elsewhere.

\subsection{PBH} \label{subsec:PBH}
\noindent
The idea of  black holes of primordial origin in the early universe was proposed in 1967 by Zeldovich and Novikov \cite{1967Nov} and also, separately, by Hawking and Carr \cite{GravCol,BhEarly, BubColl}. PBH have been a subject of great interest for dozens of years. Indeed, they can be also considered as DM candidates \cite{Carr_2020}. At the moment, constraints are imposed on a wide range of masses of a PBH as a DM candidate \cite{Carr_2021}. It has also been shown in \cite{Luca_2020, Clusters, universe6100158} that PBH could be formed in clusters, and that within this scenario constraints that have been imposed on black holes as DM candidates should be reconsidered \cite{Garcia-Bellido:2017xvr, Clusters,Calcino_2018}. 
\\
\\
Currently, several different mechanisms for the formation of PBHs have been suggested: conventional density fluctuations \cite{GravCol,BhEarly}, phase transitions of the first-order \cite{BubColl}, cosmic string collapse \cite{HAWKING1989237}, the appearance of black holes in hybrid inflation models \cite{Garc_a_Bellido_1996},  phase transitions of the second-order \cite{Rubin_2001} - one can find a review of PBHs formation mechanisms in \cite{Barack_2019}. Recently it was also shown that there is a new mechanism of formation of black holes in the multidimensional modified $f(R)$-gravity model \cite{10.3389/fspas.2022.927144}, containing tensor and Ricci scalar squared corrections. In addition to this fact, a modified $f(R)$-gravity offers solutions to many cosmological problems \cite{fRtoDM}, thus the possibilities of modified $f(R)$-gravity have been widely studied \cite{fR, ExtendedFR}. 
\\
\\
It is of great interest that the axion-like field and modified gravity could produce clusters of PBHs. They are both the case of second-order phase transition. There are papers considering clusters of PBHs as DM candidates \cite{MeNucl, MeIJMPD}. The idea is that mass distribution within the cluster and its finite size would lead to the weakening of the dynamical constraints on black holes as DM candidates. Unfortunately, the analysis performed in \cite{MeIJMPD} indicates that clusters of black holes should obey basically the same dynamical constraints as single black holes.
 
\subsection{Hadronic, hadron-like and composite dark matter}
\subsubsection{Hadronic and QCD-like composite dark matter}
\noindent
Extensions of the Standard Model are strongly motivated by its well-known theoretical and phenomenological limitations. A promising option is the addition to the set of SM fields of extra heavy fermions in confinement that interact in a vector manner with standard gauge bosons \cite{beylin1}. These scenarios of the SM extension are referred to as hypercolor (H-color) models. The symmetry group of the extended model in our case is  SU$(2n_F) \times $ G$\!\!\!\phantom{a}_ {\rm SM}$, where SU$(2n_F)$ stands for the H-color group, and $ n_F $ is the number of H-quark generations. The analysis of the group structure of H-color models, as well as the emergence of pseudo-Nambu-Goldstone bosons resulting from symmetry breaking and other peculiarities of this scenario (including the introduction of massive bound states of H-quarks in the framework of linear $\sigma-$ model), were studied in detail in the minimal version for $ n_F = 2, \, 3 $ --- see e.g. \cite{beylin2}.\\

An important feature of H-color models is the presence of stable neutral objects, which can be associated with the hidden mass candidates. In the $SU(4) \to Sp(4)$ scenario, these are the lightest neutral objects in the triplet of H-pions states and H-baryons (di-hyper-quarks). The latter ones are stable due to conservation of the H-baryon number but the neutral H-pion stability is provided by the specific symmetry leading to the conservation of a new quantum number, G-parity. Thus, in the $SU(4)\to  Sp(4)$ model, two particles are neutral and stable; in the $SU(6) \to Sp(6)$  H-color extension with an extra singlet H-quark, two more charged objects turn out to be stable, because of the symmetry with respect to two $U(1)$ groups. Such stable neutral particles with the masses $\sim 1 \, \mbox{TeV}$ are interpreted as the DM carriers resulting to necessary value of the DM abundance in the freeze-out scheme. Obviously, they are WIMPs but due to specific structure of the set of states in the hidden mass sector \cite{beylin3} some non-standard manifestations of this objects are possible, as we review below.\\

Also, we can consider some alternative type of the hidden mass originated from new massive hadrons strongly interacting with each other and with ordinary matter by gluon and/or electroweak exchanges \cite{Kuksa1}. Measurements at underground detectors (LUX, XENON1) \cite{LUX,Xenon} imposed some strict bounds on the cross sections for the interaction of these particles with nucleons.  Such new hadrons arise from a chiral symmetry model or from a model with an additional $SU(2)$ singlet heavy quark --- similar scenarios with heavy (meta-) stable hadrons are considered in the fourth-generation extensions or in ``mirror'' models. In these scenarios, genesis and prerequisites for the interpretation of new heavy hadrons --- i.e. strongly interacted massive particles (SIMPs) --- as the DM particles has been studied carefully.\\

More exactly, in the SIMP model, the hadronic hidden mass candidate is a heavy neutral hadron, $qQ$, composed from standard light quark, $q$, and new heavy quark, $Q$, which has vector interaction with photons and Z-boson but does not interact directly with W-boson. Then, contributions of new quarks to boson polarization functions (Peskin-Tackeuchi parameters) are small and the model is in agreement with the EW data. Moreover, in the simplest version we can suppose that the new heavy quark does not mix with the standard ones. Importantly, it was found that any additional neutral currents with a change of flavor are absent in this scenario.\\

This additional heavy quark participates in strong interactions with standard quarks as in the QCD. At the stage of nucleosynthesis, strongly connected bound two-quark and three-quark states of the form $(qQ)$, $(qqQ)$, $(qQQ)$ and $(QQQ)$ are formed. Main properties of new heavy mesons, namely, of the simplest two-quark states $(qQ)$, were intensively analyzed using, in particular, modified methods of heavy quarks effective theory (HQET)\cite{Hmesons}. 
It allowed to consider the excited states structure and low-energy interaction of new heavy stable hadrons with photons, leptons, and nucleons using for the lowest excited levels  an analogy with the standard heavy-light mesons. Introducing the effective vertex for the interaction of the new heavy hadrons with W-boson, cross-section of the lepton scattering on the hadron DM can be analyzed quantitatively. Also, strong low-energy interaction of new hadrons and their scattering on nucleons were carried out within the effective meson-exchange model based on dynamical realization of $SU(3)$-symmetry. Namely, the cross section of new heavy hadrons interaction with nucleons at low energies has been calculated in an analytic form and can be used to study interaction of the cold hidden mass with the matter --- clouds of hydrogen or dust, scattering of cosmic rays on the DM halo etc. --- in the Galaxy. Using the known data on the DM relic abundance, the new hadrons mass was estimated as $ \approx 10 \, \mbox{TeV}$.\\

The symmetry properties of the model entail the evaluation of fine splitting magnitude between masses of charged and neutral components in the new mesons doublet. The value of this splitting is small, it is $\sim 1\div 10 \, \mbox{MeV}$. Besides, exploring the structure of excited states it can be proved an existence of a hyperfine splitting between them with a small magnitude ($\approx 2\, \mbox{ keV}$ or slightly higher) \cite{beylin4,Hyperfine2020}. \\

The fine and hyperfine splitting can lead to the observation, in principle, of a glow (low-energy radiation) of hadronic DM, either in space or in ground-based observatories. In particular, this radiation would be more intensive an visible when generated by clumps of the hidden mass. The glow can indeed be seen in processes when the heavy DM particles scatter on the target, resulting into the appearance of excited states that radiate low-energy photons. This radiation can explain also events with $E\sim \mbox{KeV}$ at the XENON1T \cite{Xenon}, but still with low confidence level.\\

The model of SIMPs predicts also the emergence of long-lived charged states. In principle, manifestations of such heavy charged and (almost) stable states can be observed in anomalous events in the atmosphere --- heavy charged projectiles should create traces of ionized (and/or excited) atoms and molecules together with fast muons, pions, electrons and photons. More exactly, an extensive atmospheric shower (EAS) would occur that would be characterized by a specific profile. These showers can be seen presumably at low altitudes and the question is how to separate such events produced by exotic heavy mesons from analogous showers generated by the "standard" cosmic rays. 	
These scenarios, accounting for an additional heavy quark, can manifest in astrophysical and ground-based observations, as well as in collider experiments. Coming back to hyper-color scenarios inspired by technicolor and sigma model (for construction of the set of new bound states of hyper-quarks), we may note that the small mixing between the Higgs boson and an additional scalar field (an analog of known low-energy sigma meson) ensures the necessary smallness of the Peskin-Takeuchi parameters \cite{Adv2017}. This provides robustness to the aforementioned scenarios, in terms of interpretation of observed phenomena and prediction of the new ones.\\

As it was mentioned, in the hyper-color models, the hidden mass sector is multi-component and the mass splitting between the stable neutral DM candidates is $\approx 160\, \mbox{MeV}$, being nearly constant in the whole region of model parameters. The DM candidates masses have been evaluated from numerical solution of the kinetic equations system for the hidden mass burnout during evolution to the freeze-out point. With this scenario, diffuse photon spectrum generated by the DM annihilation as well as the cross sections for the DM pairs transformation into the SM particles were also calculated. Due to the lack of LHC data for the heavy hidden mass, interactions of galactic and intergalactic very high-energy cosmic rays (VHECR) with the DM are an important source of specific DM manifestations. Specifically, from the calculations of these scattering processes it has been shown that the contributions of intermediate scalars exchanges can be especially relevant \cite{beylin5}.\\

As an interesting test bench for neural networks methods in data analysis, one may suggest to consider multicomponent DM luminescence due to transitions between the mass-split components. Indeed, radiation of diffuse photons with energies up to 10 GeV, especially from regions of high hidden mass density, is possible \cite{LUM,PhysPart2023}. The effect is found in the H-color extension of the SM, where heavy neutral H-pions and H-baryons are stable and originated from different hypercolor structures \cite{Review1}. However, in any multi-component hidden mass sector analogous effects are possible, being detectable by cosmic telescopes depending on the mass splitting values. The problem is how to reliably select such photons from cosmic photon noise of various frequencies, determine bounds of energy range and detect the source of radiation and its location, if possible. 
Furthermore, the H-color scheme can be extended up to $SU(6)\rightarrow Sp(6)$, where an analogous set of heavy stable fields arise. However, there appear also two stable states with fractional charge. This urges to ponder carefully their  dynamics, manifestations and related cosmological consequences. These are open questions, available to future investigations.\\

Clearly, the hyper-color vector model does not solve all the SM problems. Nonetheless, except some interesting and characteristic effects for the multicomponent heavy DM, the model can be complemented, for instance, by clockwork scheme \cite{CW1,CW2}. This provides the necessary smallness of the neutrino masses and couplings, due to the specific hierarchy of extra fields. Immediate questions emerge how observable and measurable the manifestations of such additional fields can be. It seems reasonable that methods of data analysis that are proposed by neural networks for such multi-messenger investigations, can represent effective tools for the classification of prominent data structures, and the research of the sets of successive signals from interacting additional fields. In other words, the analysis of sequences of similar signals --- final photons with eventually diminishing energies, stable or decaying leptons that are radiated by intermediate fields, having a defined energies and etc. --- can be done more accurately by "eyes and brains" which are learned by modern neural networks.

\subsubsection{Stable multiple charged constituents of dark atoms}
\noindent
During the last decades the mainstream of studies of cosmological DM was related to the predictions of supersymmetric (SUSY) models, which had the advantage to solve the internal SM problems. The solution for the problem of divergence of the mass of the Higgs boson was among one of the most attractive features of SUSY. Simultaneously the lightest supersymmetric particle could be stable and having the mass in the range tens-hundreds GeV thus playing the role of Weakly Interacting Massive Particle (WIMP) candidate for DM --- see \cite{DMRev,Aprile:2009zzd,Feng:2010gw,4} for a review and related references.

SUSY particles have not been hitherto found at the LHC, as well as WIMPs seem not to have been detected in the underground direct searches for DM. It appeals to explore more extensive field of BSM physics and, in particular, to address possible nonsupersymmetric solutions, cutting the divergence of the Higgs boson mass. Models of composite Higgs boson can offer such an alternative solution and, as it was shown in the framework of Walking Technicolor models \cite{KK,KK1}, can also lead to a new approach to the nature of DM revealing its composite dark atom character \cite{Glashow:2005jy,BKSR1,Khlopov:2005ew}.\\

Atoms of DM, in which new stable charged particles are bound by the Coulomb force, were first proposed in \cite{Glashow:2005jy}. However, it was shown in \cite{BKSR1} that prediction of single charged stable particles inevitably leads to the overproduction of anomalous hydrogen, severely constrained in the terrestrial matter. Stable particles with charge $+1$ bind with electrons, forming atoms of anomalous hydrogen. On the other hand, stable particles with charge $-1$ bind with primordial helium in $+1$ charged ion immediately after Big Bang Nucleosynthesis, and this $+1$ charged ion also forms with electron anomalous hydrogen. Only stable particles with even negative charge $-2n$ avoid this immediate trouble, and bound with primordial helium in neutral atoms \cite{Khlopov:2005ew}. 

However, the existence of particles with charge $-2,-4...$ should be inevitably accompanied by the existence of the corresponding positively charged particles that can bind with electrons and form anomalous helium, beryllium etc, respectively. The abundance of such anomalous isotopes can be suppressed in the terrestrial matter, if the new particles have additional long range interactions making positively charged constituents of anomalous isotopes to recombine with the corresponding negatively charged particles in the terrestrial matter, thus reducing their abundance below the experimental upper limits \cite{FKS}. 

In Walking Technicolor Models, $-2n$ charged stable techniparticles can be generated in excess over their $+2n$ charged partners, equilibrated by sphaleron transitions with the baryon excess. The relationship between the excess of $-2n$ and baryon asymmetry can explain the observed ratio of baryonic and DM densities. In this case DM is in the form of dark atoms, in which the excessive $-2n$ charged particles are bound by the Coulomb force with $n$ helium nuclei. We call such dark atoms $O$He for $n=1$, being a Bohr-like atom with lepton core of $O^{--}$ with helium shell with Bohr radius equal to the size of He nucleus or $X$He for $n>1$, being the Thomson-like atom with $-2n$ leptons situated inside a nuclear droplet of $n$ He nuclei, and plan to study their structure, interaction with the matter and possible cosmological impact.\\

The Dark atom model has the advantage to explain the puzzles of direct DM searches by annual modulations of their low energy binding with Na nuclei. Strongly interacting shell of dark atoms provides their slowing down in terrestrial matter making this form of DM elusive for strategy of WIMP searches, involving significant nuclear recoil. However, dark atom interaction with nuclei can provide a low energy binding and the corresponding effect experiences annual modulations. 

This explanation \cite{iopKhlopov} is based on the following picture of $O$He interaction with nuclei. $O$He is a neutral atom in the ground state, perturbed by the Coulomb and nuclear forces of the approaching nucleus. The sign of $O$He polarization changes with the distance. At larger distances, a Stark-like effect takes place --- the nuclear Coulomb force polarizes $O$He so that the nucleus is attracted by the induced dipole moment of $O$He, while as soon as the perturbation by the nuclear force starts to dominate, the nucleus polarizes $O$He in the opposite way, so that He is situated more closely to the nucleus, resulting in the repulsive effect of the helium shell of $O$He. Qualitatively, this leads to a shallow potential well with a low energy bound state in $O$He-Na system. While such a state does not exist for heavy nuclei like Xenon. A quantitative description of $O$He-nucleus interaction with self-consistent account for the effects of nuclear and Coulomb forces is crucial for the description of the $O$He (or $X$He) evolution in stars. One can expect that dark atoms can be ionized and Supernova explosions can lead to the formation of an anomalous cosmic ray flux of stable multiple charged leptons. The search for such component and the possibility to discriminate the corresponding air-showers are challenges for the new generation of cosmic rays experiments.

\subsection{Mirror Dark Matter}
\noindent
DM may come from a parallel mirror sector of particles, with properties identical (due to parity acting between two sectors) or similar (if mirror parity is broken) to properties of ordinary matter. Mirror matter cosmology has been extensively studied in the literature \cite{ZKmirror,BlKh,BlKh1,Khlmirror} and extensively developed \cite{zurabCV,zurab} by some of us. Mirror baryons represent a case of asymmetric DM (originated due to baryon asymmetry in the mirror sector) of atomic/dissipative type, with several implications for the early Universe evolution and potentially testable in high precision data on primordial element abundances, cosmic microwave background, and distribution of matter at large and small scales. Mirror stars can exist, and first high mass mirror stars could form PBHs of astrophysical origin.  Mirror neutrinos are natural candidates for sterile neutrinos. Within the case of broken mirror parity, mirror neutrinos with keV range masses can also constitute DM. The baryon and lepton violating interactions between ordinary and mirror particles may provide a co-genesis mechanism of baryon asymmetries in both ordinary and mirror sectors, and thus naturally explain the near coincidence of the baryonic matter and DM mass fractions in the Universe. Mixing phenomena between the neutral ordinary particles and their mirror twins are of especial interest, e.g. ordinary and mirror photon kinetic mixing that can be a portal for DM direct detection, or ordinary-mirror (active-sterile) neutrino mixings. Possible mixing between ordinary and mirror neutrons can be an explanation of the neutron lifetime problem --- 10 second discrepancy between the neutron lifetimes measured via the bottle (neutron disappearance) and beam (proton appearance) experiments --- and can have interesting implications for the evolution and properties of neutron stars, for ultra-high energy cosmic rays, presence of antimatter in galactic cosmic rays, etc. In close future several experiments are planned for direct search of the neutron-mirror neutron conversions at the Paul Scherrer Institute (PSI, Villingen, Switzerland), Institut Laue-Langevin (ILL, Grenoble, France), Oak Ridge National Laboratory (ORNL, USA), and in the perspective of large scale experimental projects, neutron disappearance and regeneration experiments at the European Spallation Source (ESS, Lund, Sweden). All these aspects make mirror matter an appealing candidate which can be cross-tested via the multi-messenger approach.

The principal possibility for multi-messenger probes for mirror DM come from the correlation of signals in GW detectors from violent processes in mirror objects with observations in other tools of multi-messenger astronomy.

\section{Phenomenological Probes and Multi-messenger Constraints}\label{sec2bis}

\subsection{Constraining DM from VHECR}
\subsubsection{Interaction of VHECRs with the DM particles in the Milky Way galaxy} 
\noindent
Because high-energy cosmic rays (CR) consisting of protons, electrons, neutrinos and possibly hidden mass carriers (they can be accelerated in the process of CR scattering on the DM) have energies which are inaccessible for ground-based colliders, the SM extensions structure can be observed and quantified in a variety of astrophysical events at energy scales of about $10 \div 10^5 \, \mbox{TeV} $ and higher. Thus, careful and detailed analysis of final states for high-energy electrons, photons and neutrinos scattered on (or producing, in the inelastic scattering reaction) the DM particles can result to important conclusions about parameters and properties of new (stable) states from the SM extensions. \\

There are a number of processes of VHECR interactions with the DM demonstrating previously unknown effects and signals, which acquire particular importance due to new possibilities of cosmic telescopes and ground-based observatories like JAMES WEBB, LHAASO, HAWK, HESS and others. These new signals can be registered both separately or in correlation with neutrino events observed at the IceCube, Baikal-GVD, KM3NeT, Hyper-K etc. The analysis of these multi-particle events in such a wide energy interval is a very complex and cumbersome task. Thus the application of methods suggested by neural networks in order to detect regularities and anomalies in the data set, patterns of specific configurations of final particles, selections of specific energy intervals (where some set of events can be unambiguously explained by specific physical processes), is not only possible but also necessary. In more detail, some events can be detected at either IceCube or LHAASO or other ground-based observatories due to specific forms of rare EAS generated by fast heavy neutral objects. Such EAS should be separated from other events in the atmosphere due to studying of special classes of variables describing EAS. Note that Artificial Intelligence is already in use for the extensive air showers analysis and simulation \cite{AI_EAS}.\\
	
Using the data on EAS generated by the VHECR, it is possible to determine the nature and type of interaction of the DM components with standard particles from cosmic rays, and, therefore, to understand whether the DM objects belong to WIMPs, SIMPs or are something else \cite{beylin1,beylin2,beylin3}. It allows also to study the spatial distribution of DM near the Galaxy center, in the Galaxy halo, to search DM inhomogeneities (clumps) and to trace VHECR interaction with DM or with gas and dust clouds in the galaxy. We also suppose that the elastic and inelastic reactions of cosmic rays with the particles of hidden mass bring an  information about the DM dynamics, and that its spatial distribution may influence VHECR's energy spectra and propagation in space.\\
	
As the most informative reactions, inelastic or quasi-elastic scattering of CR particles (protons, electrons, neutrinos, photons) on DM components \cite{beylin4,beylin5} should be investigated. The products of these reactions will also be neutrinos, leptons, photons, standard (unstable) mesons, and hadronic jets. Some of these reactions have already been studied both in the framework of supersymmetry, where cosmic rays are scattered by neutralinos (wino, higgsino...), and in the hypercolor SM extension. It was shown that: the scattering cross sections for electrons and neutrinos by hyperparticles strongly depend on the DM mass; secondary particles are emitted mostly forward; channels are found that dominantly contribute into the reaction cross section; the probability of secondary neutrinos registration at the IceCube facility is small when the lepton component of CR is scattered by DM \cite{beylin5}; high-energy photons from quasar jets can effectively produce neutrinos of high energies interacting with components of hidden mass \cite{NePhot}.\\
	
It is a hard but important task to study the inelastic scattering of cosmic rays by DM particles in a halo surrounding AGNs. As a result of such VHECR scattering, not only fast secondary neutrinos arise, but also stable DM particles from halos can be accelerated to high ($\simeq 10\div100\, \mbox{TeV}$ ) energies --- these are the so called secondary VHECRs. Such neutral particles with high energies carry an information not only about the source of the primary VHECRs, but also about the nature of the interaction, properties and spatial distribution of DM objects in the halo. Signals of such reactions which are generated by interaction of secondary VHECRs with the Earth's atmosphere, can be specific in composition (corresponding EAS has a small number of muons), in the number of particles (broad showers arising from either the interaction of several neutrinos in atmosphere at different depths or at close depths). These EAS can  contain fast (heavy) neutral particles --- these are DM particles accelerated in previous scattering processes. These specific signals can be searched both among the EAS at the facilities like LHAASO \cite{LHAASO}, and in correlation with the neutrino signals at the IceCube. Ultrahigh-energy photons can also lead to the appearance of fast intergalactic neutrinos or accelerated particles DM, when photons interact with the halo DM surrounding the AGN. It is also necessary to study correlations between the neutrino fluxes detected at ground-based neutrino facilities, with photons measured by space telescopes (such as Fermi-LAT) and the composition and energy distribution of EAS recorded at the same time by cosmic rays experiments. Such correlations may indicate the reaction of deep inelastic scattering of intergalactic VHECRs on the DM particles, which simultaneously generate several secondary neutrinos and hadronic jets (in the scattering of protons from VHECR) and/or accelerated DM objects (or their excited states decaying with the formation of fast standard mesons).\\

Physical effects that occur as a consequence of the VHECR interaction with DM are diverse and informative, and their signals can be measured by different ground-based detectors. These data can help to study both the DM structures and SM extensions. These processes are studied both theoretically --- calculating analytically the cross sections of VHECR scattering on DM particles in several different scenarios --- and phenomenologically, by deploying the new generation of cosmic rays experiments. In any case, such a huge bulk of data can be effectively studied only with powerful instruments that enable the selection of "suspicious" data sets and their relative comparison, while searching for correlations based on dynamics and symmetries.  Machine learning methods provide a suitable tools for this analysis --- see e.g. Ref.~\cite{CR_ML}.

\subsubsection{Theoretical predictions for cosmic rays experiments}
\noindent
Cosmic rays  may hide invaluable indirect information on DM physics. Observation of any odd effects in CR are tried to be interpreted as possible DM signal. Positron anomaly in CR, observed in \cite{Adriani:2008zr,PhysRevLett.113.121101,2012PhRvL.108a1103A}, at energies $E\leq 1$ TeV,  possible cosmic electron-positron excess at $E\simeq 1\div 1.5$ TeV \cite{Ambrosi:2017wek} are vivid examples. 
Decay or annihilation of DM particles with production of high energy $e^{\pm}$ could account for these effects. However these top-down models of CR origin inevitably lead to gamma-production, which is hardly consistent with existing data on cosmic gammas, first of all, at highest energy part of the spectrum. Gamma yield can be changed by modifying DM decay or annihilation model, though it is not possible to fully eliminate it. So, data on cosmic gamma-radiation can provide a test ground of similar DM models, and ongoing cosmic rays experiments can significantly extend the energy range to be probed. The essential point in probing these models is that the contribution to gamma-radiation from DM with respect to astrophysical background attains maximal values at their highest energies --- close to DM particle masses, as a rule --- that are eventually inaccessible to the existing experiments. The appearance of such a growing contribution to gamma-radiation at the high energy edge, in relation to the origin of DM, provides a distinctive feature to encourage experimental investigations.

\subsection{The multi-messenger cosmological probes of new physics in cross correlation with its physical and astrophysics effects} 
\noindent
Effects of new physics in the very early Universe in addition to the necessary elements of the now standard cosmology (inflation, baryosynthesis and DM) inevitably lead to additional model dependent consequences, like primordial GW and PBHs, primordial nonlinear structures, decaying particles or early matter/dark energy dominance. The analysis of these multi-messenger cosmological probes for new physics is of special interest for super high energy scales \cite{we,DMRev,4}. 
The systematic studies of the cosmological impact of new physics was extensively developed be some of us. The approach of cosmoparticle physics then provides the chance for a cross-disciplinary study of new physics in a proper combination of its physical, astrophysical and cosmological signatures \cite{newBook}.

Cosmological, astrophysical and experimental physical probes for new physics are complementary. This complementarity of studies of physics corresponding to a new physical scale $V$ follows from the fact that laboratory effects of new physics are reduced at the energies $E \ll V$ to search for rare processes suppressed by some power of $(E/V)$, while astrophysical and cosmological effects may correspond to the high energy $E \!\gg\! V$, at which new physics can manifest in its full strength. This observation individuates in the theoretical analysis of the cosmological consequences of new physics, in cross correlation with astrophysical and experimental physical constraints, an important part of the development of the programme of studies of new physics in the new generation of cosmic rays experiments.

\subsection{Detection of decaying DM with cosmic rays experiments}
\noindent
Cosmic rays interactions in the interstellar medium provide a guaranteed source of neutrinos and gamma rays with energy range above 10 TeV. Apart from this guaranteed source, other ``unexpected" sources might appear on the sky, like the DM decay signal, which has unique spectral and imaging properties and could be readily distinguished from the the diffuse flux from cosmic ray interactions. 

Decaying DM produces both neutrinos and gamma-rays in hadronic decay channels. The cosmological contribution to the gamma-ray signal is suppressed in the gamma-ray band because of the pair production of the photons of the Extragalactic Background Light (EBL). This is not the case for the neutrino signal, which still has a cosmological (isotropic) component.

Neutrinos are detected by the IceCube experiment at the South pole. Hard neutrinos fluxes at high energies $E>200$ TeV, measured in muon neutrino channel, can be explained by conventional astrophysical sources. However, the highly diffused neutrinos flux measured by the IceCube experiment at $E<100$ TeV cannot be explained by extragalactic sources, since the corresponding diffused gamma-ray background would overproduce diffused gamma-ray measurements by the Fermi-LAT experiment. This excess neutrino flux should have Galactic origin. Also, because of the IceCube neutrino limit on signal from the Galactic plane, this excess cannot be due to usual cosmic ray interactions with the gas in the Galaxy. DM decays can explain the excess in neutrino flux measured by IceCube, but at the same time provide a large flux of galactic gamma-rays which can be detected by LHAASO \cite{Neronov:2020wir}.

In the mass range $10$~TeV~$<m_{DM}<0.3$~EeV LHAASO will provide a factor of 3-to-100 improvement of sensitivity compared to present limits from HAWC, IceCube and KASCADE. The best improvement of sensitivity is expected in the 10 PeV mass range, corresponding to 100 TeV gamma-ray sensitivity by LHAASO \cite{Neronov:2020wir}.

For decaying DM, with particle masses at the PeV scale, photons are produced inevitably even in leptophilic decaying DM models. They appear as final state radiation (FSR) and as result of inverse Compton scattering $e^{\pm}$ from DM decay off background photons.  Though, the absorption probability of these energetic photons varies with the energy. Photons scatter off different electromagnetic backgrounds (CMB, optical, radio) and that prevents them from coming from the whole Universe. This provides a non-trivial gamma-spectrum from DM decays carrying, together with signals in the neutrinos channel, a rare information on the DM physics and its distribution in space. The expected gamma-spectrum can be estimated in nearest time, in its dependence on $m_{DM}$.

\section{Gravitational-wave probes}\label{sec3}
\noindent 
The unprecedented possibility disclosed within the last decade to probe high energy physics resorting to gravitational-wave (GW) probes, can be accounted for along the following proposed lines:

\begin{itemize}
    \item {\bf GW probes for Higgs sector couplings in the SM and beyond}. GWs can be used as a source of information on triple/quartic Higgs coupling complementary to ongoing and planned measurements at colliders --- see e.g. Refs.~\cite{Huang:2016cjm,Alves:2018oct,Alves:2019igs}. One can then investigate to what extent the GW can probe possible compositeness of the Higgs sector helping to distinguish between the composite versus elementary Higgs scenarios. Since composite QCD-like scenarios typically predict rather large self-interaction couplings, these may strongly enhance the strength of the phase transitions associated with the dynamical EW symmetry breaking. Similar studies have been already developed by some of us on probing the Higgs sector properties via GWs \cite{Morais:2019fnm,Addazi:2019dqt,Vieu:2018nfq}. A cross-analysis of both GW observability limits and collider constraints becomes then mandatory. Composite models often predict scalar composite heavy DM candidates that might strongly affect the strength and other characteristics of high-scale phase transitions.

    	\item {\bf GW probes for Higgs and gauge sector couplings in the SM and beyond}. 
    	GWs from cosmological first-order phase transitions provide information on Higgs-sector interactions (including the triple/quartic Higgs couplings) that is complementary to collider measurements, see e.g. Refs.~\cite{Huang:2016cjm,Alves:2018oct,Alves:2019igs}. 
    	More generally, the GW signal is controlled by the finite-temperature scalar potential, and is therefore sensitive to Higgs-portal, gauge and Yukawa couplings in BSM scenarios; this enables quantitative GW--collider complementarity, particularly in models with dynamical electroweak symmetry breaking (such as composite Higgs setups) where enhanced self-interactions can strengthen the transition and may correlate with additional scalar/DM states.
    	Building on earlier studies \cite{Morais:2019fnm,Addazi:2019dqt,Vieu:2018nfq}, recently supercooled phase transitions in conformal $\mathrm{U(1)}^\prime$ extensions (including Majoron/seesaw realizations) and in conformal dark sectors were investigated, confronting them with current LVK/PTA data and forecasting the reach of LISA/ET \cite{Goncalves:2024lrk,Goncalves:2025uwh,Bertenstam:2025jvd}. 
    	Correlated cosmological relics were further explored, such as primordial black holes and primordial magnetic fields \cite{Balaji:2025tun}. A combined GW--collider--cosmology analysis is thus essential.

    \item {\bf GW probes for light and ultra-light DM models}
    (ALPs, specifically Majoron/Axion-like). The GW spectrum measurements provide a source of information about new physics models featuring very light pseudo-Goldstone (axion-like) scalar states that under certain conditions may play a role of DM. The presence of such states may indicate an approximate global continuous symmetry. This is for instance the case of the family symmetry at high scales that is both spontaneously and explicitly broken at low scales. Despite of a light pseudo-Goldstone mass, the presence of such states affects the vacuum structure and can lead to additional phase-transition patterns that otherwise would not exist \cite{Addazi:2019dqt}. We may pose a question about a possible complementarity of the GWs data with the direct DM detection experiments data, such as the recent hint from XENON1T measurement.
    
    \item {\bf GW probes for neutrino mass generation mechanisms}. Future GW measurement can probe the scale of Majorana neutrinos and lepton-number breaking patterns, hence zooming into the properties of the neutrino spectrum --- see e.g. Ref.~\cite{Addazi:2019dqt}.
    
    \item {\bf GW probes for the new physics energy scale in BSM scenarios}. Probing the high-scale phase transitions in LR-symmetry/family symmetry/GUT theory models --- see e.g. Refs.~\cite{Addazi:2018nzm,Camargo-Molina:2017kxd,Camargo-Molina:2016yqm} --- can be envisaged. This would offer a way to disentangle different new physics scenarios beyond the reach of particle collider measurements.

    \item {\bf GW probes for PBHs}. Future GW astronomy will open a new window into the early Universe by probing two key signatures: high-redshift black hole merger events and the stochastic GW background over a wide range of frequencies. The statistical properties and frequency content of these signals will enable us to conduct a powerful test of the existence of PBHs~\cite{Ng:2022agi,Ng:2022vbz,Franciolini:2023opt,LISACosmologyWorkingGroup:2023njw}, potentially distinguishing them from their astrophysical counterparts. 
    
\end{itemize}

All these items require building a sophisticated Machine-Learning based tool that would comprise both constraints coming from colliders measurements and analyses of the phase transitions for phenomenologically allowed points.

\subsection{GW probes for Higgs sector couplings in SM and beyond}
\noindent
A historically unique era of particle physics and cosmology has begun with the launch of operations of the Large Hadron Collider (LHC) at CERN, the largest experiment ever, as well as many large- and small-scale neutrino physics, DM, cosmic-ray and GW astrophysics experiments. SM has appeared to be the most phenomenologically successful theory of the subatomic world proven to work impressively well in interpreting all the experimental data collected at particle colliders so far. \\

It is unquestionable, however, that there are overwhelming phenomenological evidences that strongly suggest the need for a more complete theory. Indeed, the SM does not naturally incorporate a neutrino oscillations mechanism, has issues with vacuum stability and does not contain suitable DM candidates. Besides, it provides neither a reasonable explanation for the observed strong fermion mass and mixing hierarchies nor the baryon asymmetry in the Universe. In this regard, the question about the accessibility of new BSM phenomena through direct measurements becomes more and more precious. Currently, the absence of new physics indications either suggests that new particles and/or interactions can only show up at a larger energy scale beyond the current reach of collider measurements, or is due to a lack of sensitivity of the current measurements to very rare phenomena. Clearly, the greater challenge in probing such new phenomena means a weaker interplay and interaction between the SM and new physics sectors indicating a growing demand in new methods and tools.\\

A poorly known structure and dynamics of the ground state of the Universe is ultimately responsible for the majority of the unsolved problems in particle physics. It determines the particle spectra and interactions, that we observe, through the mechanism of spontaneous symmetry breaking. In particular, a specific shape of the potential of the Higgs field leads to electroweak symmetry breaking, where the ground state features a constant Higgs field value. The properties of the Higgs field potential and interactions determine the values of all particle masses and also many of the coupling strengths. This occurs through the electroweak phase transition, a few picoseconds after the Big Bang, when the current laws of physics are determined. In order to understand exactly how and why this happens we need to probe the Higgs field potential phenomenologically.\\

%The SM framework predicts a minimal well-defined structure of the Higgs field potential. Currently, we have only experimentally measured the minimum of the potatial, but we have no idea how it behaves for both large and small field values. This can be acheived by experimentally probing triple and quartic Higgs  self-interaction coupling strengths, $\lambda_{hhh}$ and $\lambda_{hhhh}$, respectively.
The SM predicts a minimal, well-defined structure for the Higgs-field potential. At present, only the potential near its minimum has been experimentally probed; its behavior at both large and small field values remains essentially unconstrained. This can be addressed by experimentally probing the strengths of the triple and quartic Higgs self-interaction couplings, $\lambda_{hhh}$ and $\lambda_{hhhh}$, respectively.
The SM predicts their specific values which are, however, highly sensitive to possible extensions of the Higgs sector. Popular BSM theories require additional scalar fields in the scalar potential that significantly affect their properties, in particular, by changing the value of $\lambda_{hhh}$ and $\lambda_{hhhh}$. Therefore, it is mandatory to measure these couplings with a sufficient precision as one of the most critical indirect tests of the SM.\\

The concepts of compositeness and supersymmetry represent the two distinct paradigms for new physics model building. The compositeness concept relies on the existence of a new strongly-coupled dynamics at large energy scales responsible for the formation of a whole family of heavy bound states composed of elementary constituents, in a close resemblance to the hadron spectra in Quantum Chromodynamics (QCD). In this approach, the Higgs boson is considered to be among the lightest composite states while the electroweak symmetry is broken dynamically via a condensation of new fermions at a new large confinement scale --- the phenomenon is in some sense similar to spontaneous chiral symmetry breaking in QCD. A typical realization of such a dynamical electroweak symmetry breaking provides a natural solution of the hierarchy problem and the vacuum stability issue in the SM. An alternative concept of supersymmetry, a new extended spacetime symmetry that relates bosons and fermions, offers another pathway towards the solution of the hierarchy problem guaranteeing the cancellation of quadratic divergences to the Higgs boson mass to all orders in perturbation theory.\\

Many particular realizations of both such concepts have been proposed in the literature so far, while the measurements cut out deeper regions of the parameter space leaving less freedom in such models and thus rendering a further progress in new physics searches more difficult. Development of new methods and tools is mandatory at this stage to increase the sensitivity to possibly weak and yet elusive signatures of new physics.

The measurement of the Higgs boson pair production at the LHC will provide a direct access to the triple-Higgs coupling value. Its possible deviation from the SM prediction represents an efficient tool for probing the possible existence of additional scalar fields as well as new fermion states entering the triple Higgs coupling through radiative corrections. Particular effort should be devoted to the study of the triple Higgs coupling, both experimentally and theoretically. Indeed one can extract crucial information about the structure of the key benchmark new physics scenarios through a comparison of the model predictions for $\lambda_{hhh}$ to the experimental data. This information would enable us to set further constraints on model parameter space and interactions in both the minimal composite Higgs model (MCHM) and Minimal Supersymmetric Standard Model (MSSM) and to potentially distinguish between these models if a statistically significant deviation of $\lambda_{hhh}$ from the SM value is experimentally observed.\\

Cosmological phase transitions (PTs) offer an inspiring possibility to probe physics beyond the SM. If first-order PTs took place at early cosmological times, a GW spectrum can be induced, with crucial observational consequences for current and future GW experiments. The recent observation of cosmic GW spectrum opened up a new channel for probing new physics in complementarity with ongoing new physics searches at particle colliders \cite{Huang:2016cjm,Alves:2018oct,Alves:2019igs}. 
Models with extended scalar sectors, typical for both compositeness and supersymmetry scenarios, feature a possibility of strong first-order electroweak phase transition (EW-PT) in the early Universe. In fact, such a possibility is not realised in the SM making it impossible to explain the origin of baryon asymmetry in the modern Universe, thus, further motivating non-minimal Higgs sectors and a more complicated structure of the scalar potential. The first-order EW-PT is capable of generating a stochastic background of primordial GWs which, if detected, can open up a gravitational portal and an efficient probe for BSM physics. The shape and evolution of the Higgs-field effective potential at finite temperatures strongly depends on microscopic structure of underlined theory. Future measurements of primordial GW spectra at planned GW spectrometers such as LISA will provide an access to the triple-Higgs coupling complimentary to that from the LHC measurements. \\

One of our key goals would be to study the potential for precision measurement of $\lambda_{hhh}$ allowing to distinguish between the supersymmetry and composite dynamics through a combination of future experimental data from both LISA and LHC. For the purpose of studies of interplay and efficient combination of collider and GW observables we plan to develop an inclusive computational framework based on state-of-the-art deep learning techniques.
We will start with an analysis of low-energy effective supersymmetric and composite theories from the perspective of triple-Higgs coupling as well as the GW observables. The focus will be on the relative comparison of MCHM and MSSM implications on $\lambda_{hhh}$ that also contain new vector like fermions (VLFs). Besides influencing the triple-Higgs coupling through radiative corrections, the VLFs might be able to generate observable signatures of primordial GWs, address the g-2 anomalies as well as provide new Higgs and exotic fermion decay signatures at colliders. Thus, we plan to perform an inclusive study of all these observables. 
The inevitable consequence of first-order PT is the formation of PBHs in bubble collisions. It provides the link between the GW spectrum and spectrum of PBHs, created in such transition, as well as implies studies of the observable effects of such PBHs.\\

In order to be able to develop these studies in an efficient way, it necessary to provide a novel computational framework based on the state of the art of deep learning techniques. At this purpose, it is possible to interface model building and Monte Carlo software tools while applying Deep Learning techniques in order to combine all available theoretical and phenomenological information. This will include an option to evaluate the impact of collider constraints or infer predictions for the primordial GW stochastic background. The final aim is to combine this knowledge with the experimental measurements of the triple-Higgs coupling to help pointing the path towards a consistent theory of the fundamental interactions.

\subsection{GW probes for PBHs}
\noindent
The direct detection of GW in 2015~\cite{LIGOVirgo2016} provided a significant boost to the study of PBHs. The idea that some of the detected black hole merger events have a primordial rather than stellar origin is extremely intriguing and widely discussed~\cite{Bird:2016dcv,Sasaki:2016jop,Clesse:2016vqa,Blinnikov:2016bxu,Ali-Haimoud:2017rtz,Raidal:2018bbj}. PBH binaries can form both in the very early Universe deep in the radiation-dominated stage due to pairwise decoupling from the Hubble flow~\cite{Nakamura:1997sm,Ioka:1998nz,Sasaki:2016jop}, and in the relatively contemporary Universe in galactic and sub-galactic structures, due to dynamical channels: two-body~\cite{Bird:2016dcv,Clesse:2016vqa} and three-body interactions~\cite{Franciolini:2022ewd}. Corresponding estimates of present-day PBHs merger rate vary by several orders of magnitude depending on the binary formation channel. The point is that the prediction of the number of PBH mergers depends in a very complex way on the fraction of PBHs in the DM $f_{PBH} = \Omega_{PBH}/\Omega_{DM}$, moreover, there is a ``mutual influence'' of the binaries formation channels on each other due to nonlinear gravitational dynamics during PBH clustering.\\

As was shown in Refs.~\cite{Afshordi:2003zb,Inman:2019wvr,Tkachev:2020uin,Amin:2025dtd}, when the contribution of PBHs to the DM density is sufficiently large $f_{PBH} \gtrsim 0.01$, natural Poisson clustering of PBHs arises and gravitationally bound dark structures form long before the first proto-galaxies. Unlike Sec.~\ref{subsec:PBH}, where the possibility of weakening constraints on PBHs through clustering was raised, this Section focuses exclusively on the observable GW manifestations of this effect. In a cluster environment, complex dynamical effects of interactions between black holes occur, which affect both the population of binaries and the internal evolution of the cluster itself~\cite{Raidal:2018bbj, Vaskonen:2019jpv,Jedamzik:2020ypm,Delos:2024poq}. In addition, Refs.~\cite{Stasenko:2023zmf, Stasenko:2024pzd,Stasenko:2025foo} showed that such effects, on the one hand, lead to a decrease in the merger rate of PBH binaries formed in the early Universe, but on the other hand, cluster dynamics leads to an increase in the population of binaries formed through dynamical channels --- see also Ref.~\cite{Franciolini:2022ewd}. Moreover, both these PBHs merger channels are comparable in the case of $f_{PBH} \gtrsim 0.1$ at relatively small cosmological distances with redshifts $z \lesssim 1$, which is within the sensitivity threshold of current ground-based GW detectors. Several theoretical models predict the initial clustering of PBHs~\cite{Rubin:2001yw,Khlopov:2004sc,Young:2015kda,Suyama:2019cst,Ding:2019tjk}. In this case, the physical distances between PBHs are much smaller on small spatial scales than on average at large scales. Furthermore, the dynamical interactions between PBHs are much stronger than in the case of Poisson clustering, significantly affecting the PBH merger rate~\cite{Bringmann:2018mxj,Young:2019gfc,Stasenko:2021vmm,Stasenko:2021wej,Eroshenko:2023bbe,Stasenko:2025mgr,Stasenko:2025vqz}. The temporal evolution of the merger rate in both Poisson and initial PBH clustering cases differs significantly from the predictions of astrophysical black hole mergers. Both models also produce distinctive spectral features in the stochastic GW background~\cite{Braglia:2021wwa,Garcia-Bellido:2021jlq,Mukherjee:2021ags,Atal:2022zux, Stasenko:2025vqz}. These features, which arise in the PBHs scenario, can be tested through future GW observations.\\

Another factor influencing the dynamics of PBH binaries, and consequently the theoretical prediction of their merger rate, is the accretion of DM halos onto PBH binaries. The binaries themselves interact with the DM particles of this halo, and this interaction leads to the evolution of their orbital parameters~\cite{Kavanagh:2018ggo,Pilipenko:2022emp,Jangra:2023mqp,Stasenko:2024dui}, in a manner similar to the GW emission. Note, that this consideration does not at all imply that PBHs do not make up all DM. PBHs with masses significantly smaller than the components of the binary system can play the role of ``light'' DM particles. in particular, the possibility that PBHs with masses $M_{PBH} \sim 10^{18} - 10^{22}$~g can make up all the DM is still open. In the latter case, the PBHs forming a halo around the binary system would be the cold component, since the interaction between such light black holes is negligible. It should also be noted that if DM poses self-interactions, the evolution of the binaries orbital parameters will generally differ due to this interaction~\cite{Berezhiani:2023vlo,Kadota:2023wlm,Fischer:2024dte,Aurrekoetxea:2024cqd,Alonso-Alvarez:2024gdz}. In any case, such interaction between the binary system and DM particles will be reflected in the GW signal from the black hole merger, which can then be used to study the nature of DM. \\

Current observations make virtually impossible to determine whether merging black holes are stellar or primordial. However, future GW detectors will possess unprecedented sensitivity, enabling to detect black hole mergers at high redshifts $z \gtrsim 10$, i.e. at a epoch when significant astrophysical black hole populations had not yet formed, and study GW across a wide range of frequencies~\cite{TianQin:2015yph,LISA:2017pwj,Reitze:2019iox,ET:2019dnz,ET:2025xjr}. This would also likely allow the orbital parameters of merging binaries to be probed, as the detectors will be sensitive to the long inspiral phase that precedes coalescence. An important point to note here is that, unlike astrophysical black holes, PBH binaries are expected to form with high eccentricities. This will allow to answer the question about the existence of PBHs more reliably, and  to identify the most promising models for their formation. Achieving this ambitious goal depends on advances in engineering, improvements to physics facilities and the development of data analysis methods, including machine learning approaches.

\section{Machine Learning applied to Particle Physics}
\noindent 
The possible adaptation of machine learning techniques to quantum field theory, and specifically particle physics, in order to attain improved performances of the codes that are used to calculate scattering processes, their associated cross-sections and the related spectra for cosmic rays, is a very fascinating and promising point of the research directions we are exploring here. In the next sections we will introduce a number of basic concepts coming from computational complexity theory, artificial intelligence and machine learning essential to draw future directions of development for DM multi-messenger astronomy.  Machine learning is not introduced here as a replacement for physical modeling, but as an inference and representation tool enabling the exploration, correlation, and compression of high-dimensional theory–data spaces inherent to multi-messenger physics.

\subsection{Computational complexity theory}
\noindent
Computation is an essential part of the technical and scientific methodology. Given a specific input to a specific set of operation models, the desired output will be produced according to the parameters of the model. The key issue consists in finding the suitable parameters that are supposed to feed the model. This captures the essence of machine learning. The complexity of the problem defines the complexity of the computation to the extent that it defines the complexity of the search space. A basic task of theoretical computer science is to sort problems into complexity classes. A complexity class contains all problems that can be solved within a given resource. Theoretical computer science identified two complexity classes of problems, respectively P (polynomial time) problems and NP (non deterministic polynomial time) problems. The conjecture that these two classes are distinct in the standard Turing model of computation is the most important open question in the field. In particular, the question can be formulated by asking whether every problem whose solution can be verified in polynomial time can also be solved in polynomial time, that is in such a way that the time to complete the task varies as a polynomial function on the size of the input to the algorithm (as opposed to, say, exponential time).\\

As a generality, each physical theory supports computational models whose power is limited by the physical theory. It is well known that classical physics supports a multitude of implementations of the Turing machine. Non-Abelian topological quantum field theories exhibit the mathematical features necessary to support a model capable of solving all \#P problems (sharp P-complete problems), a computationally intractable class, in polynomial time. Specifically, Witten \cite{witten:1989quantum} has identified expectation values in a certain SU(2)-field theory with values of the Jones polynomial \cite{jones:1997polynomial} that are \#P-hard \cite{jaeger:1990computational}. This suggests that some physical system whose effective Lagrangian contains a non-Abelian topological term might be manipulated to serve as an analog computer capable of solving NP or even \#P-hard problems in polynomial time. Defining such a system and addressing the accuracy issues inherent in preparation and measurement is a major unsolved problem. A polynomial-time algorithm for solving a \#P-complete problem, if it existed, would solve the P versus NP problem by implying that P and NP are equal. No such algorithm is known, nor is a proof known that such an algorithm does not exist. \\

In 1993 computer scientists Ethan Bernstein and Umesh Vazirani defined a new complexity class called BQP, which is an acronym that stands for ``bounded-error quantum polynomial time'' \cite{bernstein:1997quantum}. More specifically, BQP is the class of problems solvable with high probability in polynomial time by a quantum computer. Around the same time, Bernstein and Vazirani also proved that quantum computers can solve all the problems that classical computers can solve. That is, BQP contains all the problems that are in P. Moreover, BQP contains problems not found in another important class of problems known as PH, a different acronym that stands for ``polynomial hierarchy'' \cite{raz:2019oracle}. PH is a generalization of NP. In particular, PH contains almost all well-known complexity classes such as P, NP, and co-NP. This means PH contains all the problems if one starts with a problem in NP and make it more complex by layering qualifying statements like ``there exists'' and ``for all''. Classical computers today cannot solve most of the problems in PH, but theoretical computer science assumes that PH is the class of all problems classical computers could solve if P turned out to equal NP. In other words, to compare BQP and PH is to determine whether quantum computers have an advantage over classical computers that would survive even if classical computers could (unexpectedly) solve many more problems than they can today.\\

Neto et al. have proved that Quantum Neural Network models coupled with a non-unitary operator can solve the 3-SAT problem in polynomial time, a result that no classical neural network was able to obtain previously \cite{Neto:2015SolvingNP}. As we will discuss in chapter \ref{QNN}, one of the approaches one can pursue consists in making use of (Topological) Quantum Neural Networks for building models of particle physics. This approach allows to deal with the complexity of the computational space associated with this problem by exploiting the computational resources coming from the quantum computing platform and, possibly, therefore solving problems unsolvable by using a classical platform. In this sense, it should be obvious why the computational capabilities of the underlying platform could play an important role in shifting not only the computational efficiency of the learning algorithm but also the class of problems that could be eventually faced from an algorithmic point of view.

\subsection{Machine learning: perspectives and approaches}
\noindent
A machine learning model is a mathematical function $Y = f(X)$ between an input X and an output Y. The goal of different machine learning approaches is to learn this function given some observed input and output data. It is useful to consider machine learning problems from several perspectives:
\begin{itemize}

\item   
{\bf Machine learning as probabilistic inference.} A first perspective is that machine learning tasks are often tasks involving probabilistic inference of the learned model from the training data and prior probabilities. In fact, the two primary principles for deriving learning algorithms are the probabilistic principles of Maximum Likelihood Estimation --- in which the learner seeks the hypothesis that makes the observed training data most probable --- and Maximum a Posteriori Probability (MAP) estimation --- in which the learner seeks the most probable hypothesis, given the training data plus a prior probability distribution over possible hypotheses. The perspective that machine learning algorithms are performing probabilistic inference is very compatible with the perspective we are going to mention according to which machine learning algorithms are solving an optimization problem. In most cases, deriving a learning algorithm based on the MLE or MAP principle involves first defining an objective function in terms of the parameters of the hypotheses and the training data, then applying an optimization algorithm to solve for the hypothesis parameter values that maximize or minimize this objective.
  
\item {\bf Machine learning as optimization.} Machine learning tasks are often formulated as optimization problems. For example, in training a neural network containing millions of parameters, we typically frame the learning task as one of discovering the parameter values that optimize a particular objective function such as minimizing the sum of squared errors in the network outputs compared to the desired outputs given by training examples. When machine learning tasks are framed as optimization problems, the learning algorithm is often itself an optimization algorithm. Sometimes we use general purpose optimization methods such as gradient descent (e.g., to train neural networks) or quadratic programming (e.g., to train Support Vector Machines). In other cases, we can derive and use more efficient methods for the specific learning task at hand (e.g., methods to calculate the maximum likelihood estimates of parameters for a na\"ive Bayes classifier).
 
\item 
{\bf Machine learning as parametric programming.} Another perspective we can take on the same learning programs is that they are choosing parameter values that define a function or a computer program written in a programming language which is defined by their hypothesis space. For example, we can view deep neural networks as implementing parameterized programs, where the learned network parameters instantiate a specific program out of a set of potential programs predefined by the given network structure. As we move from simple feedforward networks, to networks with recurrent (feedback) structure, and with trainable memory units, the set of representable (and potentially learnable) programs grows in complexity.
  
\end{itemize}

As concerns the main approaches, the literature distinguishes between parametric, non-parametric, and semi-parametric models.
\begin{itemize}
    
\item A parametric approach assumes the functional form (i.e. the shape) of the mathematical function \(f \) by construction. That is, it assumes that the function belongs to a particular family of mathematical functions such as, for instance, linear, quadratic, etc. The goal is now to determine the coefficients (parameters) of the different components of the function on the basis of the training data. More specifically, a parametric model will assume some finite set of parameters $ \theta $. Given the parameters, $\theta$, future predictions X will be independent of the observed data $ D: P(X |\theta,D)= P(X|\theta) $. Therefore, $\theta$ captures everything there is to know about the observed data D. As a consequence, the complexity of the model is bounded even if the amount of data is unbounded. Yet, parametric models suffer from being poorly flexible to the extent the shape of the function is defined a priori. Linear regression is an example of such an approach. It is assumed that the input and output follow the relation $Y= \beta_0  + \beta_1 X$ and the goal is to determine the values of the parameters, i.e., the coefficients $\beta_0$ and $\beta_1$. Note that, in this case, the function belongs to the family of linear functions and different values of the parameters will generate different linear functions. The main task of a parametric approach is equivalent to estimating the vector of parameters. Logistic regression, k-means and hidden Markov models are all examples of parametric models.
    
   \item Non-parametric models assume that the data distribution cannot be defined in terms of such a finite set of parameters. But they can often be defined by assuming an infinite dimensional $\theta$. Usually we think of $\theta$ as a function. The amount of information that $\theta$ can capture about the observed data D can grow as the amount of data grows. This makes them more flexible. In this sense, a non-parametric approach does not make any assumption about the functional form, it is very flexible and can take any shape. It could be a very complex function, combination of extremely large number of non-linear functions or it could be a rule like large margin boundary or it could be a simple estimation of density or discriminant outcome in the desired input space. Gaussian approaches, k-nearest neighbour and decision trees are examples of non-parametric approaches. The term non-parametric does not mean that such models are completely lack parameters, but that the number of the parameters are flexible and not fixed priori.
    
    \item Finally, semi-parametric modeling is a hybrid of the parametric and non-parametric approaches of statistical models. It may appear at first that semi-parametric models include non-parametric models; however, semi-parametric models are considered to be “smaller” than a completely non-parametric model because we are often interested only in the finite dimensional component of $\beta$. By contrast, in non-parametric models, the primary interest is in estimating the infinite dimensional parameter. In result, the estimation is statistically harder in non-parametric models compared to semi-parametric models. While parametric models are being easy to understand and easy to work with, they fail to give a fair representation of what is happening in the real world. Semi-parametric models allow to have the best of both worlds: a model that is understandable and offering a fair representation of the messiness that is involved in real life. Semi-parametric regression models take many different structures. One is a form of regression analysis in which a part of the predictors does not take pre-determined forms and the other part takes known forms with the response. For example, $f_1$ may be known (assume linear) and  $f_2$ is unknown. In this case, the semi-parametric form will be written as $Y= \beta_0  + \beta_1 x_1+f(x_2 )$. In this setting, the relationship between $x_1$ and the response is linear but the relationship between the response and $x_2$ is unknown. The most illustrative example is the scatter plot smoother. The approach is model-based that utilized the basic principle like Maximum Likelihood Estimation (MLE). Mixed model-based smoother can be extended to a full hierarchical Bayesian model when analyzed via Markov chain Monte Carlo \cite{ruppert:2003semiparametric}.
\end{itemize}

Direct implementation of neural network models is based on the use of common architectures and their modifications, such as the multilayer perceptron (MLP), single-layer and multilayer $n$-dimensional convolutional neural networks (CNNs), recurrent neural networks (RNNs), usually included through the modern LSTM or GRU cell models. MLPs are commonly implemented via dense or fully-connected pre-defined layers in popular APIs such as Keras, PyTorch or MATLAB Deep Learning Toolbox, and can be used on their own to solve a wide range of problems, or as an integral part of hybrid models. 
Convolutional neural networks excel in image processing tasks such as spectrogram and telescope data processing, strong lensing detection, and feature extraction, and have recently been widely used in GW data processing. Recurrent neural networks demonstrate high efficiency in time series processing tasks, for example, in processing gamma-ray bursts data and identifying patterns in changes of observed stars' characteristics. Transformer neural networks, as well as graph neural networks (GNNs)\cite{NIPS2016:6418}, are rapidly gaining popularity in solving problems in various fields, including computer vision and time-series analysis. Each of the mentioned architectures can be part of a more complex organizational structure, such as a Variational Auto-Encoder (VAE) \cite{dai:2018syntax, Ciardiello:2020dtr, tf4z-71rt} or a Generative Adversarial Network (GAN) \cite{Lin:2019htn}.
More general approaches to machine learning include supervised and unsupervised methods such as Support Vector Machines (SVMs), Random forests and gradient boosting algorithms, for example, XGBoost\cite{Aad:2019vvf}, clustering methods(KMeans and others), K-nearest neighbors algorithm.
With the introduction of physics informed machine learning (PIML) \cite{mlic1,mlic2,mlic3,mlic4,mlic5} and neural networks (PINNs) \cite{smith2025pinngrapephysicsinformedneural,Mishra_2025,BENTO2025139690}, the possibility was opened to combine the flexibility and versatility of neural networks with the accuracy and ability of physical interpretation of partial differential equations. This approach allows to find numerical data-driven solutions to problems formulated in terms of nonlinear PDEs. This results is achieved by taking into account physical constraints, patterns and domain knowledge through the use of a neural network as a solver, and by encoding information about the structure and solution of the differential equation in the memory of the neural network during training. The same approach also enables to solve the problem of discovery of differential equations based on neural network calculations, both for discrete-time models and continuous-time models, as well as for problems with dynamic and periodic boundary conditions.

\subsection{Explainable AI and Knowledge Discovery}
\noindent
In science, the formal definition of a theory is difficult but commonly it refers to a comprehensive explanation of a subfield of nature that is supported by a body of evidence \cite{humphreys:2016oxford}. In physics, the term theory is generally associated with a mathematical framework derived from a set of basic axioms which allows to generate experimentally testable predictions for such a subfield of physics. Typically, these systems are highly idealized, in that the theories describe only certain aspects of the reality. Verifiability of a theory has been progressively substituted by falsifiability \cite{popper:19341959}. Generally speaking, the use of machine learning algorithms increases the level of abstraction realized by physical theories as well as the explainability power of physical theories. Machine learning methods have been remarkably successful for a wide range of application areas in the extraction of essential information from data. An exciting and relatively recent development is the uptake of machine learning in the natural sciences, where the major goal is to obtain novel scientific insights and discoveries from observational data. A prerequisite for obtaining a scientific outcome is domain knowledge, which is needed to gain explainability, but also to enhance scientific consistency. Explainable machine learning in view of applications in the natural sciences appears to play a major role. Three core elements are relevant in this context: transparency, interpretability, and explainability.

Transparency connotes some sense of understanding the mechanism by which the model works. A machine learning approach is transparent if the processes that extract model parameters from training data and generate labels from testing data can be described and motivated by the design approach. For instance, we can obviously describe as stochastic gradient descent works. Generally, to expect a machine learning method to be completely transparent in all aspects is rather unrealistic; usually there will be different degrees of transparency. For instance, in the case of neural networks, the choice of hyper-parameters such as learning rate, batch size, etc., has a more heuristic, non-transparent algorithmic nature. The same remark can be applied to the number of units or layers of the topology of the network. Due to the presence of several local minima, the solution is usually not easily reproducible as well; therefore, the obtained specific solution is not fully algorithmically transparent.

Interpretability means to present some of the properties of machine learning models in terms understandable to a human. Ideally, one should be able to answer to a question such as ‘‘can we understand on what the machine learning algorithm bases its decision?’’ Note that, in contrast to transparency, to achieve interpretability the data is always involved. \\

Explainability deals with three classes of questions: (1) what–questions, such as ‘‘What event happened?’’; (2) how–questions, such as ‘‘How did that event happen?’’; and (3) why–questions, such as ‘‘Why did that event happen?’’. The goal of the machine learning ‘‘user’’ is very relevant as concerns explainability. 
There are essentially four reasons to seek explanations: to justify decisions, to enhance control, to improve models, and to discover new knowledge. It is important to differentiate between scientific explanations and algorithmic explanations. \\

As concerns natural sciences, a broad framework leverages unsupervised learning approaches to learn low-complexity representations of physical process observations \cite{roscher:2020explainable}. A syntax-direct variational autoencoder (SD-VAE) has been recently introduced where syntax and semantic constraints are used in a generative model for structured data. As an application, the drug properties of molecules are predicted \cite{dai:2018syntax}. The learned latent space is visually interpreted, while the diversity of the generated molecules is interpreted using domain expertise. In a similar way, the classical unsupervised algorithms like principal component analysis and the derived interpretable principal components have been used for the exploration of different phases, phase-transition, and crossovers in classical spin models \cite{wang:2016discovering}. Implementations of symbolic systems based on semantic technologies suitable to improve explanations for non-insiders have been adopted. The paper \cite{YUAN2025102078} proposes a symbolic regression approach to the study of Gray-Body Factors (GBFs), which play a crucial role in the derivation of Hawking radiation and are recognized for their computational complexity. The authors explore simple analytical forms for the GBFs of the Schwarzschild black hole. To achieve this, authors employed the open-source PySR Python library as the basis for their pipeline - ReGrayssion. While the traditional approach involves determining unknown coefficients or distributions based on observed or synthetic data, symbolic regression results in analytical expressions obtained by machine learning, artificial intelligence, and heuristic algorithms. This approach allows us to identify hidden patterns in observational data that would be lost through purely numerical calculations and that are not detected by traditional fitting methods. Symbolic regression offers a unique opportunity to obtain physically justified and observationally supported expressions that are not directly derived analytically, but also retain the possibility of interpretation. The machine learning related components such as input features, hidden layers, and computational units, and predicted output of deep learning models can be mapped into entities of Knowledge Graphs or concepts and relationships of ontologies (knowledge matching). Traditionally, these ontology artifacts are the results of conceptualizations and practices adopted by experts from various disciplines \cite{futia:2020integration}.

\subsection{Collider physics application}

\noindent
Along the multi-messenger perspective that motivates this review, at least two main lines of investigation can be identified for the consistent transposition of machine-learning toolkits either borrowed from artificial intelligence or developed ad hoc, into fundamental physics applications. These two directions concern, respectively, data related to the observation of very high-energy cosmic and gamma rays, and data associated with gravitational wave observations.
In the former case, particular attention can be devoted to the analysis of very high-energy gamma-ray and cosmic-ray spectra measured by the new generation of cosmic-ray experiments  --- for a general overview across the energy and intensity frontiers see e.g.~\cite{mlic3}. In this context, data-analysis strategies and computational techniques originally developed for collider physics can be naturally adapted and transferred. This represents one of the main objectives of the present programme. Many of these techniques are already available within the machine learning framework and can be applied to air-shower reconstruction with minimal modifications, while others may require further methodological development. \\

Concerning applications inspired by collider physics, we identify the following points of immediate relevance for data analysis and dark-matter phenomenology:

\begin{enumerate}

\item

Reliable models have been developed to simulate nuclear interactions in contexts such as ion therapy, including the Boltzmann–Langevin One Body (BLOB) model or efforts in QCD matter at extreme conditions \cite{mlic4}, which describes heavy-ion interactions up to a few hundreds of MeV {\cite{Napolitani:2014waa}}. In this approach, the final state is represented as a probability density function (PDF) describing the likelihood of finding a nucleon at a given point in phase space. Due to the large computational cost of BLOB simulations, deep-learning-based emulation strategies have been developed {\cite{Ciardiello:2020dtr}}. These methods rely on the discretization of the PDF and the training of a Variational Auto-Encoder to reproduce it. In particular, Ref.~{\cite{Ciardiello:2020dtr}} demonstrates the successful emulation of BLOB PDFs for ${}^{12}$C–${}^{12}$C interactions at 62 MeV, achieving distributions consistent with the original simulations at negligible computational cost. Furthermore, enhanced control over the generation process has been obtained by reorganizing the VAE latent space to explicitly encode the dependence on the impact parameter, together with the training of a dedicated classifier. As discussed in the work programme section, this strategy is not restricted to the energy range for which it was originally developed and can be generalized to other physical scenarios.

\item 
A second major development is the deployment of machine learning algorithms at the trigger level, including implementations on FPGAs and heterogeneous hardware. At the LHC and future colliders, ML-enhanced triggers enable real-time event selection under severe latency and bandwidth constraints, allowing rare or unconventional signatures to be retained already at the earliest stages of data acquisition. Techniques such as quantized neural networks, graph-based inference, and low-latency autoencoders have been successfully ported to FPGA architectures {\cite{duarte:2018a}},{\cite{francescato:2021a}}, demonstrating that sophisticated ML models can operate reliably in real-time experimental environments. This development is of direct relevance for multi-messenger experiments, where similar constraints arise in radio telescopes, gravitational-wave interferometers, and extensive air-shower arrays, and where rapid identification of candidate events is essential for coordinated follow-up observations.

\item

Another important class of applications concerns model-agnostic searches for new physics, where the goal is to detect deviations from the Standard Model without committing to specific signal hypotheses. Recently proposed approaches {\cite{Belis:2024a}} based on anomaly-detection frameworks exploit unsupervised or weakly supervised learning to identify statistically significant excesses in suitably learned feature spaces, rather than in predefined kinematic variables. These methods are particularly powerful in scenarios involving composite dark matter, hidden sectors, or long-lived particles, where the signal morphology may not be well captured by traditional analyses. Crucially, the same philosophy can be applied to multi-messenger data sets, for instance to search for unexpected spectral or temporal features in cosmic-ray fluxes, neutrino events, or stochastic gravitational-wave backgrounds, without relying on detailed signal templates.

\item

A further application concerns the assessment of hadronic interaction models used in air-shower reconstruction from very-high-energy cosmic-ray observations, namely SIBYLL 2.3c {\cite{Engel:2019dsg}}, QGSJet II-04 {\cite{Ostapchenko:2010vb}}, and EPOS-LHC {\cite{Pierog:2013ria}}. A systematic scan of the corresponding output spaces can be performed using graph neural networks, which have already proven effective in the classification and regression of hadronic jets at the LHC. Compared to CNN-based approaches, GNNs naturally overcome the requirement of data structured on regular 2D or 3D grids, and can be applied directly to generic three-dimensional point clouds without imposing geometric symmetries. Similar techniques have been used at ATLAS to analyze atypical hadronic jets, such as those arising from displaced decays of long-lived neutral particles in hidden-sector or hidden-valley models. This approach appears particularly well suited for experiments such as LHAASO and can be further optimized for real-time and trigger-level applications on high-speed processors. Methods based on explainable AI can also be integrated to optimize model performance and parameter tuning, enabling a comparative assessment of different hadronic models within a unified framework.

\item 

Assessment of the three hadronic models hitherto applied to air-shower reconstructions from VHECR observations, i.e. Sybil 2.3c {\cite{Engel:2019dsg}}, QGSJet II-04 {\cite{Ostapchenko:2010vb}} and EPOS-LHC {\cite{Pierog:2013ria}}, and consequently a scan of the output space, can be achieved within this perspective applying peep neural networks based on graph-neural network already used for classification problems and regression of hadronic showers (jet) of several different types at LHC. This procedure would allow to solve the problem, related to convolutional neural networks (CNN) implemented in deep learning visual applications, of obtaining data structured in symmetric meshes (2D and 3D pixels). Furthermore, the same procedure could be applied to any 3D point clouds, without any specified request of geometric symmetry associated to the detectors. Atypical hadronic jets at ATLAS have been already analyzed adopting these techniques, this encodes for instance hadronic jets produced by the displaced decays of long lifetime neutral particles in hidden models, including either the hidden sector or the ``hidden valley’'. This situation seems particularly suitable to be adapted to the case of LHAASO, and one can naturally adapt these algorithms to high speed processors for trigger/real time analyses applications. Methods connected to explainable AI can be also applied to this purpose, in a different way than standard AI models. Therefore, an optimization strategy can be achieved, while making use of a parameter tuning, through the synthesis of the best performing tCNN adapted to the trigger FPGA, as applied to air-shower reconstruction, making use of the three different hadronic models previously highlighted.

\item

A general and essential feature of machine-learning applications to collider physics is the ability to handle high-dimensional ensembles of data and to perform fast, real-time processing. A broad review of deep learning applications to LHC physics can be found in \cite{mlic2}. Techniques developed for collider experiments can therefore be directly transferred to the reconstruction of air showers, combining information from Cherenkov detectors, as well as electronic and muonic channels. Deep-learning methods are particularly effective in extracting physical quantities from large data sets when detailed analytical models are unavailable, due to complex detector geometries or large experimental uncertainties. In this respect, deep neural networks act as universal approximators, capable of learning hierarchical representations of the data. Each layer of the network performs a transformation that progressively builds more abstract and informative representations, enabling the extraction of relevant physical features directly from experimental observations.

\end{enumerate}

In summary, collider-driven ML techniques foster a shift from traditional, channel-by-channel analyses toward a representation-based view of experimental data, where events are embedded into latent spaces that capture their essential physical content. Physics informed ML has also become quite an attractive proposition for collider physics --- see e.g.~\cite{mlic1,mlic2,mlic3,mlic4,mlic5}. Physics informed machine learning (PINNs and related approaches) has become an established direction \cite{mlic1}.
Altogether, these collider-inspired machine-learning strategies provide the computational and conceptual tools needed to efficiently explore high-dimensional parameter spaces and to coherently integrate collider, cosmic-ray, and gravitational-wave data within a unified multi-messenger inference framework for dark matter and physics beyond the Standard Model.

\subsection{Illustrative multi-messenger machine learning pipeline}
\noindent
The purpose of this section is to provide an illustrative example that should be interpreted as a conceptual demonstration of the proposed framework rather than a full quantitative implementation.\\

We consider a simplified proof-of-concept scenario in which three representative classes of observables are combined:
\begin{enumerate}
    \item gamma-ray spectra associated with dark matter annihilation or decay,
    \item high-energy neutrino fluxes compatible with IceCube-like sensitivities,
    \item stochastic gravitational-wave spectra generated by first-order phase transitions.
\end{enumerate}

Each observational channel is encoded through a dedicated feature extractor (e.g. convolutional or fully connected neural networks), producing latent representations that are subsequently combined within a joint inference architecture. The resulting model maps multi-messenger inputs to underlying physical parameters, such as the dark matter mass, interaction cross section, and phase-transition strength.
\\

We stress that this construction is intended as a conceptual template illustrating how heterogeneous datasets can be consistently integrated. In particular, no detector-level simulations, background modeling, or systematic uncertainties are included at this stage.
\\

A complete quantitative implementation would require:
\begin{itemize}
    \item realistic mock data generation for each experimental channel,
    \item detector response modeling,
    \item training and validation on statistically meaningful datasets,
    \item evaluation of parameter reconstruction accuracy and uncertainty calibration.
\end{itemize}

Such an analysis constitutes a substantial dedicated study and is beyond the scope of the present review. Instead, the goal here is to highlight the architectural principles and to clarify how multi-messenger information can be combined within a machine-learning framework.
\\

This schematic example serves to bridge the conceptual discussion of previous sections with the research directions outlined in the concluding road-map.

\subsection{Spanning the Parameter Space of Dark Matter Models with ML}
\noindent 
The collider-based applications discussed above illustrate how machine learning enables the extraction of physically meaningful information from high-dimensional data and the identification of non-standard signatures in complex experimental environments. However, the ultimate challenge in the multi-messenger program is not limited to analyzing individual experiments in isolation, but rather to consistently explore and constrain the vast parameter spaces of dark-matter and BSM scenarios across multiple observational channels. In this broader context, machine learning becomes a key instrument for spanning, organizing, and correlating theory spaces with heterogeneous data sets. Here we focus on ML-based strategies for efficient parameter-space exploration and global inference, emphasizing their role in unifying collider, gravitational-wave, cosmic-ray, and cosmological probes within a single coherent framework.\\

\noindent
The Standard Model provides an extremely successful description of particle interactions, yet it does not account for dark matter, neutrino masses, or the baryon asymmetry. We refer to the Introduction for a general discussion of these motivations. The search for new physics is a great challenge of contemporary science, which nonetheless has not been hitherto successful in recovering detectable signatures that could be distinguished from noise backgrounds. Within this framework, machine learning techniques have been recently deployed to overcome related phenomenological problems. Probably the discovery of the Higgs has provided the most striking chance for the exploitation of techniques borrowed from artificial intelligence. Nonetheless, there exist still several impeding issues, related to the statistical limitations of the samples employed for the training, the way of dealing with uncertainties, the reliability of the results and the occurrence of model-dependent interpretations. Within this wide scenario one can identify a plethora of possible searches that can be beneficial to artificial intelligence methods, and that can be used as for concrete demonstrations of toolkits. These include mainly DM candidates searches, originated from the decay of new particles, as predicted by supersymmetry, supergravity and other models that we have been reviewing in the preceding sections. Among these possible searches, we can also consider specific parameter space regions of either ALPs particles or ``dark photons'', as light particles that belong to novel hidden sectors not yet observed because too feebly interacting with ordinary matter.\\

Within this wide perspective, one may deploy machine learning techniques that can be adapted to span the large parameter spaces of the different models hitherto proposed. Parameters spaces can be so extended that concretely assessing specific regions might require a sizable computational time, hence affecting the concrete capability to accomplish phenomenological analyses. Machine learning techniques allow to avoid these restrictions. Developing both theoretically and numerically the phenomenology of the DM models to be assessed in light of the multi-messenger perspective is an urgent point to unveil the nature of the dark universe.

\subsubsection{The multi-messenger approach and cross-correlation with other channels} 
\noindent
The multi-messenger strategy requires to process jointly information arising from the gravitational radiation channel, the electromagnetic radiation channel, the cosmic rays and neutrino channel and the terrestrial collider channel. This strategy involves methods of cosmoparticle physics, which includes system of analysis of multi-messenger cosmological probes for new physics, as well as indirect searches for DM, and new frameworks developed to make contact between deep neural network and quantum field theories. This theoretical treatment is aimed to specify in the analysis of the data of multi-messenger astronomy possible parameters of new physics and make in their terms predictions for observable features in cosmic rays experiments. The methodology we are going to specify at this purpose will make heavily use of machine learning techniques, which has become a distinctive and essential feature of phenomenological analyses. \\

Training samples needed to the implementation of the toolkits that are proper of deep learning are constituted by theoretical predictions. This choice enables a simulation that may allow to scan the parameter spaces of several different DM models. The machine learning techniques hence deployed, and further developed to this purpose, can help restoring the spaces to be scanned, increasing the numerical performances of the phenomenological analysis, while decreasing the required computational time. 
The different methodologies, deployed along the different observation channels that can be selected, can be then finally recomposed in a unified analytical framework, so to limit the parameter spaces of the theories to be assessed. This cross-fertilizing strategy may then enhance the sensitivity of the scrutinized models to the observations provided the relevant experiments.

\subsubsection{Interaction Networks, First Order Phase Transitions and BSM models}
\noindent
We are currently living in the data/information era, with new trends emerging from the creation to storing to finally reading and interpreting the information conveyed by the data itself. From the scientific community perspective, this trend started to emerge from the last decade, because of the rapid investment on many types of experiment/sections. The fast paced evolution of hardware and the data availability within international co-operations are the driving force of such shifting trend. Encoding information and reasoning about objects, relations and physics is one of the main domains humans takes for granted, and its among the most basic and important aspects of intelligence. Nearly all of the physics we experience in our daily basis can be described in terms of the interactions rules (symmetries, forces, space-time transformations, ...) between their components (particles, fields, objects, ...). \\

A recent work \cite{Cranmer:2020wew} demonstrated a way to leverage the inductive biases from of the Interaction Networks (INs) \cite{NIPS2016:6418} in a Graph Network, to learn models of particle systems at different domains. In this respect, a similar approach can be deployed to tackle the rather challenging problem of using the combined information from particle collisions, cosmic rays and future space-based GWs detectors as a way to probe and constrain BSM theories.

To have a better understanding of the challenges we will face it is first necessary to lay down the basic concepts of GWs generated by the early universe. The GWs produced during the earlier stages of the universe are a consequence of the cosmological first-order phase transition (FOPT), in which the universe changes from a meta-stable high energy (symmetric) phase to a stable lower energy (broken) phase \cite{Coleman:1977py}. This can occurs via quantum tunnelling or thermal energy transfer, and a consequence from this change is the appearance of bubbles of the broken phase, separated from the surrounding symmetrical phase by a thin wall. These bubbles expand at relativistic velocities and pushing the surrounding plasma, eventually some bubbles can also collide or coalesce. This process generate large amount of energy which can be converted in GWs \cite{Witten:1984rs,Kosowsky:1991ua} and the characteristic power-spectrum of these GWs are directly tied with the theory model.  

We still do not fully comprehend all the steps which our Universe took to transit from the earlier symmetric to the current broken phase, and knowing these steps is crucial for a better understanding of the nature behaviour from low to higher energy scales. There is a plethora of BSM models, each one capable to explain and predict phenomena associated with particular energy scales --- see e.g. Refs.~\cite{Morais:2019fnm,Addazi:2019dqt,Vieu:2018nfq}. These models are often parametrized in terms of effective field operators, which provides a scheme that offers an easy and powerful method to understand the interactions between the fields involved in a given theory. The effective operators are composed of the interactions fields (scalars, fermions, gauge, etc) and a constant that encodes the information and the energy scale at which the interaction occurs. 

One important aspect of such approach, is that we can structure build any BSM model as an expansion of effective operators, where these new operators can modify the behaviour of the phase transition and consequently the GWs power spectrum. This aspect can potentially be used as a way to track the effects of these operators on the GW. To perform such a task it is crucial to leverage the inductive biases emerged from the Interactions Networks in their generalized form, the Graph Network. A main objective is to use Graph Networks as a way to learn from simulated data --- these can be produced by means of packages such as CosmoTransition \cite{Wainwright:2011kj} and/or PhaseTracer \cite{Athron:2020sbe} --- the subtle effects in the GW spectrum that can arise from the interactions terms predicted by the BSM. This may allow indeed to derive a symbolic expression bridging the model parameters to the GW data, with a consequent better understanding of the individual roles of the BSM models at the FOPT during the earlier stages of the Universe. One can then build a robust analysis for the GW from FOPT and a new framework based on machine learning capable of predicting a GW spectrum for a given theory as well as constrain BSM theories and get new insights from simulations in order to prepare the theoretical background for analyzing the data of future space-based experiments.

\subsection{Adapting models of particle physics to quantum neural networks}
\label{QNN}
\noindent
BSM models of particle physics can be developed by casting them directly on the graph configurations of quantum neural networks. This strategy should be implemented proceeding along the directions outlined by Refs.~\cite{QNN, QNN2, QNN3}, which specify how a topological quantum field theory (TQFT) can be associated to a class of neural networks that represent a quantum version of deep neural networks in machine learning.     
In TQFT indeed, one considers principle bundles of the relevant gauge Lie group, and hence takes into account: i) the holonomies of the Lie algebra connection that is associated to the principle bundle: these are the new configuration variables of the theory; ii) the fluxes of the frame field, which are the conjugated variable to the holonomies. This theoretical construction is also known as a $BF$ theory, serving as a prototype of TQFT. There exist a precise mathematical path one can resort to in order to recast Yang-Mills theory directly in term of TQFT, as their deformations. This traces back to the formulation of those ``extended'' (namely, not anymore topological) $BF$ theories that are constructed on the principle bundle of the Lie group, the Yang-Mills theory of which we aim at representing --- see e.g. \cite{Cattaneo:1997eh}. \\

This construction is not the only one provided hitherto in the literature. Pretty recently it has been suggested in Ref.~\cite{Cranmer:2020wew} that a discretization strategy may be applied to systems with finite matter degrees of freedom. The smearing process is supported by graphs that are at the base of the neural networks construction. This strategy is then extended so to encode scalar matter degrees of freedom, not only vector gauge fields. In other words, a discretization procedure can be accounted for, either resorting to the association path summarized in \cite{Cranmer:2020wew} or to discretization method pushed forward in \cite{QNN}. Within the former case, an analogy among graph network theory and Newtonian mechanics allows to associate nodes to particles, pair of nodes to pair of interacting particles, edge among nodes to forces acting on the latter, and so on. Within this latter case, a detailed discretization procedure can be derived from lattice formulations of field theories \cite{Creutz:1984mg}, including vector fields and scalar fields. Furthermore, fermionic fields can be encoded in the picture, discretizing degrees of freedom at the nodes of the graphs.\\

Within this scenario, the very novelty of the approach proposed in Refs.~\cite{QNN,QNN2,QNN3} is that dynamics is implemented on the 2-complexes associated to the neural network structures. Hidden layers are supported on graphs, rather than on nodes, as in previous attempts in the literature \cite{Beer}. This allows to provide a functorial dynamical evolution of the states of the theory, which are supported on 1-complexes, i.e. graphs, and is provided by the discretized path-integral on the lattice (scalar, gauge and fermionic) theories considered. Developing such a framework immediately provides some advantages. First, a non-perturbative treatment is naturally implemented within this context, which allows to accommodate wider inspection into the parameter spaces of the theory, without the necessity to account for perturbativity constraints anymore. Furthermore, issues related to gauge invariance, which are often encountered while resorting to cut-off regulators to perform loop-integral evaluations, are easily avoided. These issues can actually affect in a relevant manner the estimated observables of the theory, and thus in general provide a source of ambiguity and a limit of the predictive power of the model. Finally, the discretization on the lattice of quantum fields is well tailored to achieve a natural implementation of the algorithms developed from the side of machine learning, at least the ones for which state of the Hilbert space are supported on the graphs of the quantum neural networks.
Hence, a proper span of the parameter space that characterize models that arise as SM extensions, can be in principle be better accommodated by the development of non-perturbative methods. Nonetheless, the latter can be linked through quantum neural networks to machine learning techniques, making feasible the task of optimizing the parameters of the model \cite{QNN3}.

\section{Cosmological inferences from Machine Learning}
\begin{figure}[!h]
    \centering
    \includegraphics[width=1\linewidth]{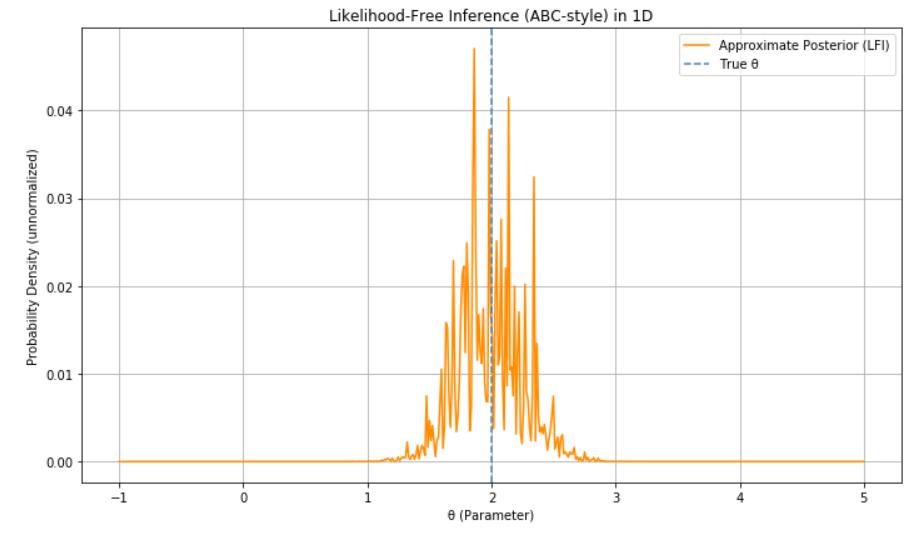}
    \caption{Demonstration of likelihood-free inference (LFI) using a simulation-based approximate posterior in a 1D toy cosmological parameter estimation task. Unlike traditional Bayesian inference, which requires an analytical likelihood, this approach uses distance-weighted simulation outputs to recover the posterior around the true parameter value. Such methods are foundational in machine learning-based cosmological inference pipelines where forward simulations are available but explicit likelihoods are intractable.}
    \label{1}
\end{figure}

\begin{figure}[!h]
    \centering
    \includegraphics[width=1\linewidth]{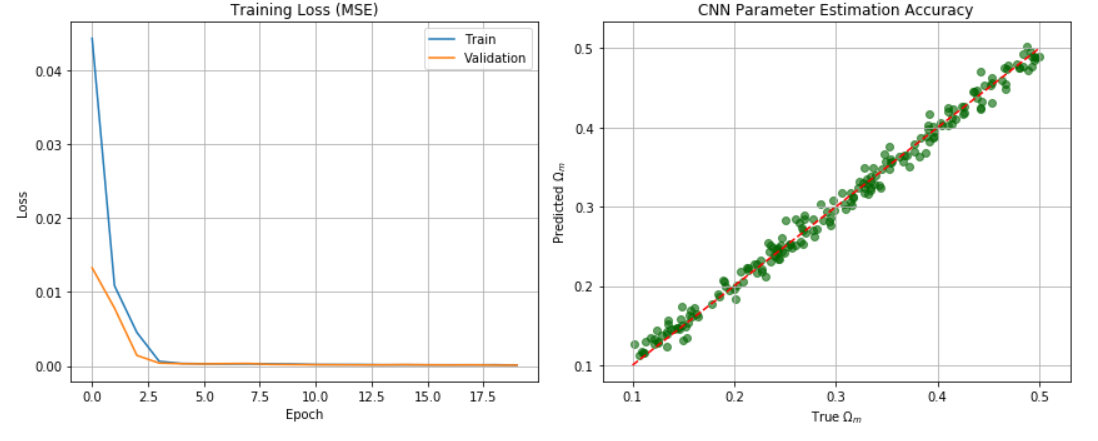}
    \caption{Convolutional Neural Network (CNN) trained to estimate the cosmological parameter $\Omega_m$ directly from synthetic 2D "sky map" images. The left panel shows the training and validation mean squared error (MSE) loss over epochs, indicating successful learning. The right panel compares predicted and true $\Omega_m$ values on held-out data, demonstrating the model’s strong regression performance. This illustrates the potential of CNN-based inference in cosmology.}
    \label{2}
\end{figure}

\noindent
The complexity and scale of modern cosmological data, particularly from surveys like Euclid \cite{euclid}, LSST \cite{LSST:2008ijt}, and DESI \cite{DESI:2025zgx}, require inference frameworks that are not only computationally efficient but also capable of managing non-Gaussian, high-dimensional, and noisy data distributions. Machine learning and more specifically deep learning thus emerges as a robust framework capable of enhancing or replacing traditional inference methods, especially when likelihood functions are analytically intractable or prohibitively expensive to evaluate \cite{Ho:2019zap1979,Peel:2018aei1980,Caldeira:2018ojb1981,He:2018ggn1982,Ravanbakhsh:2016xpe1983,Escamilla-Rivera:2019hqt1984,ANTARES:2018uvq1985,Lanusse:2017vha1986,cosmoverse}.
\\
In Bayesian inference, one seeks the posterior distribution
\begin{equation}
    p(\boldsymbol{\theta}|\boldsymbol{d}) = \frac{L(\boldsymbol{d}|\boldsymbol{\theta})\pi(\boldsymbol{\theta})}{Z(\boldsymbol{d})},
\end{equation}
where $L(\boldsymbol{d}|\boldsymbol{\theta})$ is the likelihood, $\pi(\boldsymbol{\theta})$ the prior, and $Z(\boldsymbol{d})$ the evidence. However when the likelihood is not analytically tractable, methods such as Approximate Bayesian Computation (ABC) or simulation-based inference become necessary.
\\
Neural networks, particularly artificial neural networks (ANNs), can be trained to emulate forward models such that the mapping $\boldsymbol{\theta} \mapsto \boldsymbol{d}$ can be learned. Let $f_{\text{NN}}: \mathbb{R}^n \to \mathbb{R}^m$ represent the trained neural mapping from a parameter space to a simulated data space. Then inference proceeds via inverse learning
\begin{equation}
    \boldsymbol{\theta}_{\text{MAP}} = \arg\max_{\boldsymbol{\theta}} \, p(\boldsymbol{\theta}|f_{\text{NN}}^{-1}(\boldsymbol{d}_{\text{obs}})).
\end{equation}
This approach, known as likelihood-free inference (LFI) \cite{Wang:2020sxllike1,Wang:2023vejlike3,Gomez-Vargas:2021zyllike2,Manrique-Yus:2019hqclike4}, enables posterior sampling via emulators such as the AGB (Artificial Gaussian Bayesian) \cite{SpurioMancini:2021ppkagb,Albers:2019rztagb2,Auld:2006pmagb3,Wang:2020hmnagb4} networks used in recent cosmological analyses. We illustrate likelihood-free inference in figure \ref{1} by reconstructing a posterior distribution using only simulations and a distance-based kernel, without requiring a closed-form likelihood.
Bayesian Neural Networks (BNN) offer probabilistic predictions by placing a distribution over the weights $\boldsymbol{w}$ of the network. Given data $\boldsymbol{D}$, the predictive distribution for a new input $x^\star$ is
\begin{equation}
    p(y^\star|x^\star, \boldsymbol{D}) = \int p(y^\star|x^\star, \boldsymbol{w}) p(\boldsymbol{w}|\boldsymbol{D}) \, d\boldsymbol{w}.
\end{equation}
This integral is typically approximated using variational inference or Monte Carlo dropout. BNNs have been employed for redshift estimation, modified gravity model selection, and classification of theoretical cosmological models from power spectra \cite{Gunther:2022pto96}. They provide not only point estimates but also credible intervals, crucial for cosmological parameter constraints.
\\
One can also consider Convolutional Neural Networks, which are employed to directly map image-based cosmological observables (e.g., weak lensing maps, 21cm intensity cubes) to parameters. For input tensors $X \in \mathbb{R}^{H \times W \times C}$, CNNs use convolutional layers defined as:
\begin{equation}
    (X * K)_{i,j} = \sum_{m,n} X_{i+m,j+n} \cdot K_{m,n},
\end{equation}
where $K$ is the convolution kernel. This allows feature extraction without the need for explicit summary statistics. Applications include axion mass constraints, GW parameter estimation, and weak lensing analyses \cite{Sabiu:2021aeacnn,asorey2012recoveringcnn2,Andres-Carcasona:2023rnkcnn3}. In figure \ref{2} we simulate synthetic 2D sky-like maps with statistical features tied to a cosmological parameter, $\Omega_m$, and train a CNN to regress that parameter.
 The left panel shows rapid convergence of training and validation loss, indicating stable learning. The right panel plots true vs predicted $\Omega_m$, demonstrating the CNN's high accuracy and generalization.
This exemplifies how deep learning enables direct cosmological inference from image-like observables, bypassing handcrafted summary statistics.
\\
One also employs Normalizing Flows (NFs) \cite{nfpapamakarios2021normalizing} models which take in complex, high-dimensional distributions by transforming a simple base distribution $p_z(z)$ (usually Gaussian) via a series of invertible and differentiable mappings $f_i$
\begin{equation}
    \boldsymbol{\theta} = f_n \circ \dots \circ f_1(z), \quad z \sim \mathcal{N}(0, I).
\end{equation}
The transformed distribution $p(\boldsymbol{\theta})$ is then obtained via the change-of-variables formula:
\begin{equation}
    p(\boldsymbol{\theta}) = p_z(z) \left| \det \left( \frac{\partial z}{\partial \boldsymbol{\theta}} \right) \right|.
\end{equation}
This formulation allows flexible modeling of cosmological posteriors. Frameworks like EMUFLOW use NFs to combine constraints from multiple datasets and mitigate dimensionality issues in joint posterior modeling \cite{nf2Bevins:2022qsc,nf3Friedman:2022lds,nf4Mootoovaloo:2024sao,nf5Friedman:2022lds}.
\\
Another possible avenue of significant interest in the cosmological regards is that of diffusion models \cite{dmsohl2015deep,dm2song2020denoising}. These models construct posteriors by reversing a stochastic noise process. Starting from white noise, they iteratively "denoise" to generate samples from a target distribution \cite{dm2song2020denoising}. Let $x_T$ be a noise-corrupted sample and $x_0$ the original data, then the forward process is
\begin{equation}
    q(x_t|x_{t-1}) = \mathcal{N}(x_t; \sqrt{1 - \beta_t} x_{t-1}, \beta_t \mathbf{I}),
\end{equation}
and the generative model learns the reverse transitions $p(x_{t-1}|x_t)$. These models are well-suited for large-scale structure inference and CMB simulations, with high fidelity and tractable uncertainty quantification \cite{dm3Mudur:2023smm,dm4Mudur:2024fkh}. What we have hitherto discussed does not encompass any novel method with respect to previously existing ones, but is rather supposed to be an overview of the practices covered in the literature so far for ML approaches to Cosmology.\\

Reference~\cite {NewGravML2024tre} is devoted to the development of a hierarchical pipeline for classification using deep learning. The authors discuss the new enhancements of ResNet-based deep learning code - AresGW, and employ the modified model to detect new GW candidate events in data from a network of interferometric detectors. The original model consists of 27 residual blocks, comprising two convolutional layers. Five blocks are used to decrease the dimensionality, and two blocks are used to reduce the number of channels and reduce the problem to a binary classification problem. The improved model includes a set of hierarchical conditional filters for different frequencies. The proposed development made it possible to reduce the number of false alarms by at least 70\% in the Default Low-Pass triggers class, and by 90\% in the Selective Noise Rejection triggers class, compared to the number of false alarms in the Default Low-Pass triggers class alone. The proposed algorithm made it possible to identify 10 events with a probability $p_{astro}\ge 0.99$ on a test data, and prior events $GW190916\_200658$, $GW200305\_084739$, $GW190906\_054335$, $GW200106\_134123$, with probabilities 1.00, 1.00, 0,99, 0.95, respectively.\\

Detailed accuracy estimates are provided in Ref.~\cite{j3mm-zjsv}. The authors propose a high-precision supervised interpolation model using AutoGluon to infer ALP DM parameters from reconstructed NS mass–radius curves. The developed model provided $R^2>0.998$ for axion-like particles mass estimations, with MAE being equal to 3.1661, RMSE = 10.1552, MAPE = 2.0339 \%. The estimates for DM Fermi momentum were 0.9992, $3.73\times10^{-5}, 3.99\times10^{-4}, 0.3399 \%$ for $R^2$, MAE, RMSE, MAPE, respectively.
An hybrid approach was proposed in \cite{tf4z-71rt}, where the authors use VAE to learn low-dimensional latent embeddings of three  synthetic datasets for black hole masses and spins. Extracted data are processed using a set of machine learning algorithms, specifically, KMeans and Random forest. The supervised random forest algorithm provided a accuracy estimates within the 0.8-0.95 range, while unsupervised KMeans approach achieved only 0.51-0.56. \\

Reference \cite{smith2025pinngrapephysicsinformedneural} then proposed physics informed neural network based on CNN architecture that can infer the mass parameters of binary black hole merger systems to within 7\% from only the time-frequency representation of the GW signal. The problem of modeling gravitational collapse, a critical phenomenon in astrophysics and cosmology, is discussed in \cite{Mishra_2025}, where a MLP-based PINN is proposed. The authors introduce two Schrödinger–Poisson informed neural networks that solve the nonlinear equations to simulate the gravitational collapse of fuzzy dark matter (FDM) in both 1D and 3D settings. \\

\begin{figure}[!h]
    \centering
    \includegraphics[width=1\linewidth]{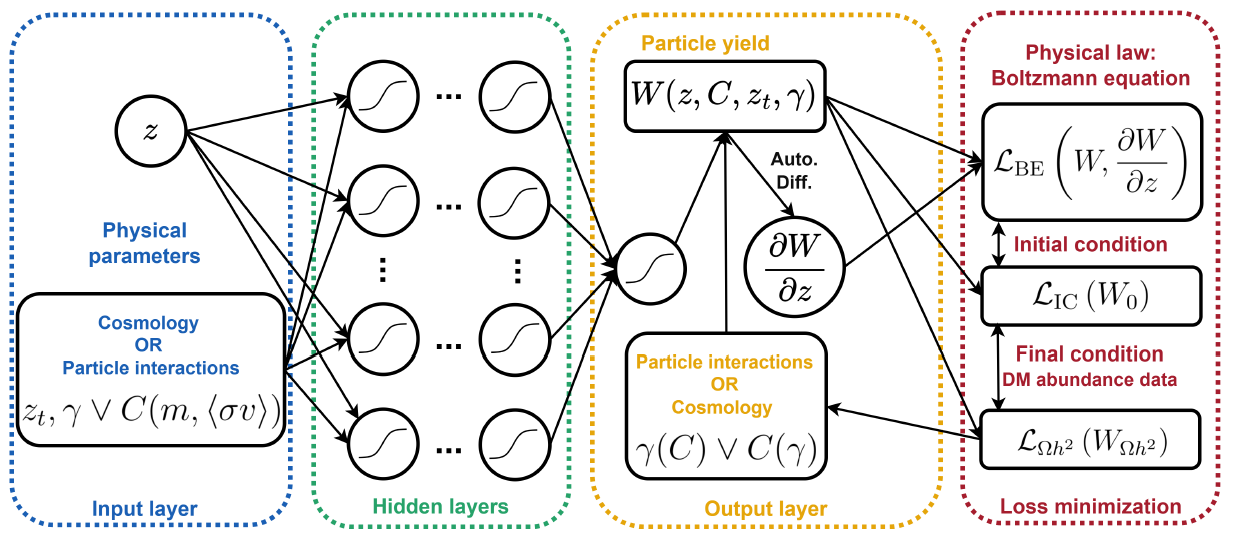}
    \caption{Figure 6, from \cite{BENTO2025139690}, shows a schematic representation of the PINN structure. This is relevant to tackle inverse problems while modeling the Boltzmann equation for freeze-in DM particle yield, in alternative cosmology.}
    \label{fig:PINN}
\end{figure}

Reference \cite{hs9b-drwx} finally provided a comparison between autoencoder-based Bayesian inference algorithm for the GWB measurement and cross-correlation method. The proposed MSMH autoencoder achieved good sensitivity much faster than the cross-correlation method, being able to estimate simultaneously a weaker GWB component. The approach allowed authors to measure a CBC GWB of amplitude $10^{-9}$ and a cosmological GWB of amplitude $1.3\times10^{-10}$. Multiple models for forward and inverse modeling of Boltzmann equation for freeze-in DM particle yield in alternative cosmology are provided in \cite{BENTO2025139690}. The authors employed multiple dense layers in a PINN architectures, restricted by Boltzmann equation and initial conditions, particle interactions and cosmological constraints to derive the estimates for pair-wise relationship between power-law exponent and particle interactions. Additionally, Bayesian methods are used to quantify the epistemic uncertainty of theoretical parameters found in inverse problems. The most promising results are obtained for inverse problems. Figure \ref{fig:PINN} from \cite{BENTO2025139690} (originally, figure number 6 there) illustrates the approach.
\\
In general, the successful deployment of ML in cosmology can introduce a new paradigm in which
\begin{itemize}
    \item Posterior estimation is data-driven, relying on accurate training simulations $\{(\boldsymbol{\theta}_i, \boldsymbol{d}_i)\}$;
    \item Forward models (Boltzmann solvers, N-body simulations) are replaced or accelerated by differentiable emulators;
    \item Likelihood-free inference replaces Gaussian assumptions with flexible neural mappings.
\end{itemize}
However, the mathematical rigor of these methods still faces challenges:
\begin{enumerate}
    \item {\bf Calibration of Uncertainties}: ML posteriors must be statistically calibrated to avoid overconfident predictions.
    \item {\bf Explainability}: Black-box networks can obscure physical interpretability. Work on physics-informed architectures and interpretable flows is ongoing.
    \item {\bf Generalization}: Trained networks must generalize across cosmologies; adversarial training or Bayesian ensembles may help.
    \item {\bf Posterior consistency}: For ML-based posteriors $p_\text{ML}(\boldsymbol{\theta})$, we require convergence in probability to the true posterior as $N \to \infty$ (in sample size), which remains an open theoretical problem.
\end{enumerate}

The authors of \cite{NewGravML2024tre} note the difficulty of solving the binary classification problem (distinguishing a GW signal from a noise signal) using classical methods based on minimizing the mean squared error or maximizing the logarithmic likelihood function (as in minimizing cross-entropy). Events of interest in GW analysis are rare, necessitating working with unbalanced datasets. Classical loss functions in these situations can exhibit insufficient classification accuracy and suppress the detection of events in the tails of the distribution. However, machine learning approaches can mitigate these issues by using alternative loss functions, such as the Kullback-Leibler divergence, as well as hierarchical and multi-stage methods and models.
Another important aspect is that many machine learning models, in particular those based on convolutional neural networks and transformer neural networks, allow for the efficient use of parallel computing to obtain results significantly faster than traditional approaches based on pure Bayesian inference. For example, the model, proposed in \cite{Andres-Carcasona:2023rnkcnn3}, allowed the generation of 10,000 posterior samples in just 0.05 seconds using CNN, running on an NVIDIA GeForce RTX 2080 Ti GPU. The proposed approach allows to reduce time costs by several orders of magnitude compared to traditional methods, while maintaining the comparable accuracy, with only a single outlier, and reducing the likelihood of effects caused by the instrument bottleneck during the inference. Another example is \cite{blfk-7k9f}. Here the authors introduce a technique to enhance the reliability of GW parameter estimation using attention maps, obtained from spectrograms, in order to provide the model with an ability to focus on the most important data sections. The total estimation was performed in six minutes on an NVIDIA GeForce RTX 3090 using a simulated data for GW150914, proving that machine learning can significantly reduce the computation time. Reference \cite{FastDeepLearn2025} proposes a hybrid transformer deep learning network for PBH populations probing using GW data, that could be run on a single GPU. The accuracy is comparable to that obtained via hierarchical Bayesian inference. The proposed model's inference takes $\mathcal O(1)$ second on a GPU and 10 seconds on a CPU. The authors provide an estimate for a MCMC sampler, that takes about one week to converge.\\

It is also worth noting that the development of likelihood-free inference (LFI) and simulation-based inference (SBI) has progressed rapidly in recent years, becoming a central methodological framework in modern cosmological data analysis. Systematic reviews of SBI techniques have highlighted how neural density estimators, neural ratio estimation and neural posterior estimation have significantly improved the efficiency and scalability of inference pipelines for high-dimensional datasets \cite{sbi1delaunoy2024low,sbi2cranmer2020frontier}, like the ones relevant for cosmology \cite{com1alsing2019fast}. Similarly, Bayesian neural networks and probabilistic deep learning approaches have matured to the point where they can provide calibrated uncertainties and principled posterior approximations, allowing their application to problems such as cosmological parameter estimation, model selection and large-scale structure reconstruction. There are comprehensive discussions of these developments, including methodological comparisons in recent reviews and methodological papers. \cite{com2jeffrey2025dark,com3wang2023sbi,com4mediato2025cosmological,com5zhong2026machine,com6rojas2025systematic}.

\section{Future perspectives}
\noindent
The directions outlined in this work can be organized into a concrete research program:\\
\begin{enumerate}
\item
development of joint multi-messenger pipelines combining data from LHAASO, IceCube, Fermi-LAT and gravitational-wave observatories;\\
\item
implementation of machine learning frameworks for global parameter inference across heterogeneous datasets;\\
\item 
systematic exploration of benchmark dark matter models within this unified framework;\\
\item
application of anomaly-detection techniques for model-independent searches;\\
\item
investigation of next-generation methods, including physics-informed and quantum-enhanced machine learning.\\
\end{enumerate}

These directions outline a possible path toward transforming multi-messenger astrophysics.\\

In conclusion of this review of theoretical and numerical methods, we mention a few urgent points to be developed in the multi-messenger investigation on DM.\\
  
First of all, from the side of the DM models to be constrained, specific features of several types of SM extensions must be carefully considered, focusing on the additional fermionic degrees of freedom and effects that are associated with the presence and possible manifestation of DM particles in various reactions. Several points shall be taken into account.

\begin{itemize}
    
\item 
Qualitative and quantitative analyses of the processes of high-energy neutrinos (with energies in the range from 10 to 1000 TeV) scattering on the DM particles --- with the production in the final state of both electronic and muon neutrinos and/or the acceleration of the target (so called, scattered-up reactions) --- can be provided. In the scatterings, charged components of the hyperpion triplet can be produced with subsequent decay. The possibility of registering the products of such reactions at the LHAASO and IceCube is then a direction to be investigated. 

\item
In-depth study of the processes of high-energy protons and photons interactions with the DM in the Galaxy halo can be carried out. These reactions can result in the production of charged partners of the DM particles, and their decays can generate neutral stable particles together with high-energy fluxes of electrons, positrons and neutrinos. These fast secondary particles can produce specific EAS with a low muon content accompanied by neutrinos of various flavors and/or heavy neutral stable particle.

\item 
The dependence of the secondary neutrino and lepton fluxes on the spatial distribution of the DM in the Galaxy can be considered in detail. In particular, one can study the scattering of cosmic rays by inhomogeneities in the Galactic halo. This investigation can be considered in connection with the problem of studying the internal dynamics of DM in the Galaxy.

\item
Since the hypercolor model has a whole set of H-hadrons that are located higher in mass than the stable DM particles, there should be also excited unstable states of di-hyperquarks. They also can manifest themselves in scattering reactions. To clarify the SM extension type, the study of the unstable H-hadrons and the analysis of their excitation channels and decay modes are very important. Signals of such decays will also be fluxes of ordinary (decaying) mesons, accompanied possibly by stable neutral heavy DM objects. Consequently, one can expect the appearance of EAS products, rare and specific in composition and angular distribution which can be detected at the LHAASO, HAWK and HESS facilities.

\item
For a complete mass spectrum study in the H-color model, it is beneficial to consider vacuum condensates structures in the SM extensions with additional fermions. In a sense, the model contains a kind of QCD duplication but with a smaller number of quark flavors. The symmetry violation also requires the introduction of a H-quark vacuum condensate along with a nonzero vacuum condensate of H-gluons. Thus, it becomes possible to study the H-hadron characteristics using a previously unknown analogue of the QCD sum rules. At the same time, the presence of a certain hierarchy in the structure of vacuum condensates of H-color models, as well as the stability of their vacuum state, can be also investigated. Issues related to the conditions for the applicability of these methods, possibility to extract information from new types of sum rules, and to the of data on the masses of new heavy states, shall be investigated within this framework.

\item 
Interaction processes (inelastic and quasi-elastic scattering) of high-energy particles of dark and ordinary matter can be investigated, and detailed analyses of the possibilities to register the heavy metastable hadrons signals can be carried out: the conditions and channels for the production of new heavy hadrons, the types and intensity of their annihilation signals shall be considered in detail.

\item
Essential is also to carry out a study of the DM halo interaction with the gas-dusty and solid components of the Galaxy. Such an analysis may allow to obtain (after a quantitative consideration and classification of types of signals) additional information on the possible detection of DM particles. 

\item
The basis for a detailed study of the conditions and specificity of the luminosity of hadronic DM is an analysis of the hyperfine splitting between excited states. This is crucial to characterize the possible observed manifestations of DM both in the hadronic and in the hypercolor scenario. At this purpose, the character of the splitting and the conditions for the metastability of excited states of new hadrons are equally important for various scenarios of the SM extensions considered.

\end{itemize}

Concerning the numerical toolkits exploited to carry out phenomenological analyses, specific techniques derived from ML can be applied, in a multi-messenger perspective, to achieve both air-showers reconstruction of cosmic rays detection, and to scan the parameter spaces of DM models from the future measurements of the GW stochastic background, as produced by FOPTs.  We list below a few of the most relevant methods that can be implemented, consistently with community-level directions summarized in \cite{mlic5}.

\begin{itemize}
    
\item 

Enhanced VAE techniques currently under development allow to structure the latent space and control it much better than in previous studies \cite{AI1, AI2, AI3}. Elements of the latent space hence individuated will be orthogonal to the different latent variables, hence allowing for more efficient simulations. These techniques can be applied to multiple contexts, without being limited to the low-energy regime investigated so far. Instead, they belong to a framework of generative deep learning that is useful to simulate any kind of process. While it is possible to use a deep learning approach to emulate a model developed to simulate nuclear reactions in any energy regime, toolkits implementing the generation part in C++ must be still developed, and then interfaced with common Monte Carlo toolkit such as Geant4.

\item 

Tests can be carried out employing gradient boosted decision trees (XGBoost), Gaussian processes based on Bayesian regression, deep neural networks (DNN) and convolutional neural networks (CNN). Improved sensitivities by a factor from 2 to 5 in XGBoost can be achieved within several models — see e.g. Ref.~\cite{Aad:2019vvf}.  These preliminary results can be consolidated with a systematic comparison of the methods implemented, and with related tests of their robustness against input variations. In particular, in order to overcome possible biases, generative adversarial networks \cite{Lin:2019htn} will be used to study the dependencies on simulated Monte Carlo (MC) samples of the training processes.  Concerning the issues related to explainable AI in air-shower reconstructions, studies may be developed that are based on convolutional neural network (CNN) for image classification that are trained to distinguish background processes from signals that can be used to map clusters of hadrons (jets) in 3D.  This methodology has been already checked against overtraining, ensuring that an accuracy above 90\% can be achieved. This is actually susceptible to factor 2-3 improvements with respect to previous methods \cite{ATLAS2020}. Along this direction one can also adapt Graph Neural Network (GNN) techniques \cite{GNN}, hence enhancing sensitivities to new physics through machine learning techniques, while maintaining the transparency of the involved section processes. Thus, the implementation of explainability becomes here crucial to ensure the correct physical interpretation and the scientific consistency of results that are derived.

\item
A novel computational framework based on the state of the art of deep learning  techniques can be applied to the of study of cosmological phase transitions induced by thermal corrections that may leave gravitational footprints in the form of a stochastic GW background. Such a GW background may carry imprints of phase transitions above the electroweak scale, offering access to physics beyond the reach of collider experiments. It is possible to interface model building and Monte Carlo software tools while applying Deep Learning techniques in order to combine all available theoretical and phenomenological information. This will include an option to evaluate the impact of collider constraints or infer predictions for the primordial GW stochastic background. This knowledge can be combined, in this way, with the experimental measurements of the triple-Higgs coupling, to help pointing the path towards a consistent theory of the fundamental interactions.
    
\end{itemize}

\section*{Acknowledgment}
\noindent 
The research of A.K., T. B., M. K., D. S.  and V.K. was carried out in the Southern Federal University with financial support from the Ministry of Science and Higher Education of the Russian Federation (State contract FENW-2026-0028). The work of V.S. was funded by the Ministry of Science and Higher Education of the Russian Federation, Project "Studying physical phenomena in the micro- and macro-world to develop future technologies" FSWU-2026-0010. 

\bibliography{references}

\end{document}